\documentclass[superscriptaddress,prd,twocolumn]{revtex4}
\usepackage{latexsym}
\usepackage{amsthm}
\usepackage{amsmath}
\usepackage{graphicx}
\usepackage{amssymb}
\usepackage{hyperref}
\usepackage{bbm}

\newcommand{\gsim}{\lower.7ex\hbox{$\;\stackrel{\textstyle>}{\sim}\;$}}
\newcommand{\lsim}{\lower.7ex\hbox{$\;\stackrel{\textstyle<}{\sim}\;$}}

\def\epm{e^+e^-}

\newcommand{\GeV}{\,\mathrm{GeV}}
\newcommand{\MeV}{\,\mathrm{MeV}}

\newcommand{\be}{\begin{equation}}
\newcommand{\ee}{\end{equation}}
\newcommand{\bea}{\begin{eqnarray}}
\newcommand{\eea}{\end{eqnarray}}
\newcommand{\f}[2]{\ensuremath{\frac{#1}{#2}}}
\newcommand{\bef}{\begin{figure}[htbp]\begin{center}}
\newcommand{\eef}{\end{center}\end{figure}}

\newcommand{\halfPage}[1]{\begin{minipage}{0.49\textwidth}
\begin{center} #1 \end{center}\end{minipage}}

\begin{document}

\pagestyle{plain}

\title{
\begin{flushright}
\mbox{\normalsize SLAC-PUB-13650}\\
\mbox{\normalsize SU-ITP-09/22}
\end{flushright}
%\vskip 15 pt

New Fixed-Target Experiments to Search for Dark Gauge Forces}
\author{James D.~Bjorken}
\affiliation{Theory Group, SLAC National Accelerator Laboratory, Menlo Park, CA 94025}
\author{Rouven Essig}
\affiliation{Theory Group, SLAC National Accelerator Laboratory, Menlo Park, CA 94025}
\author{Philip Schuster}
\affiliation{Theory Group, SLAC National Accelerator Laboratory, Menlo Park, CA 94025}
\author{Natalia Toro}
\affiliation{Theory Group, Stanford University, Stanford, CA 94305}
\date{\today}
\begin{abstract}

Fixed-target experiments are ideally suited for discovering new
MeV--GeV mass $U(1)$ gauge bosons through their kinetic mixing with
the photon.  In this paper, we identify the production and decay
properties of new light gauge bosons that dictate fixed-target search
strategies.  We summarize existing limits and suggest five new
experimental approaches that we anticipate can cover most of the
natural parameter space, using currently operating GeV-energy beams and
well-established detection methods.  Such experiments are particularly
timely in light of recent terrestrial and astrophysical anomalies
(PAMELA, FERMI, DAMA/LIBRA, etc.) consistent with dark matter charged under
a new gauge force.

\end{abstract}
%\keywords{Beam Dumps, Very Cool Experiments...etc}
\maketitle

%%%%%%%%%%%%%%%%%%%%%%%%%%%%%%%%%%%%%%%
\section{New Gauge Forces} 

The interactions of ordinary matter establish that three gauge forces
survive to low energies.  Two striking features of these forces ---
electroweak symmetry-breaking at a scale far below the Planck scale 
and apparent unification assuming low-energy supersymmetry --- have
driven model-building for a quarter-century.  But the strong and
electroweak forces need not be the \textbf{only} ones propagating at
long distances.  Additional forces, under which ordinary matter is
neutral, would have gone largely unnoticed because gauge symmetry
prohibits renormalizable interactions between Standard Model fermions
and the other ``dark'' gauge bosons or matter charged under them.

There is an important exception to the above claim:  
new ``dark'' Abelian forces can couple to 
Standard Model hypercharge through the kinetic mixing operator $\f{\epsilon}{2}
F^Y_{\mu\nu} F'^{\mu\nu}$, where $F'_{\mu\nu} =
\partial_{[\mu}A'_{\nu]}$ and $A'$ is the dark gauge field
\cite{Holdom:1985ag}. If the $A'$ is massive, 
Standard Model matter acquires milli-charges proportional to $\epsilon$ under the massive $A'$.
Kinetic mixing with $\epsilon\sim 10^{-8}-10^{-2}$ can be generated at any
scale by loops of heavy fields charged under both $U(1)'$ and
$U(1)_Y$, and the $A'$ can acquire mass through
a technicolor or Higgs mechanism. 
A mass scale near but beneath the weak scale 
is particularly well-motived ---
$U(1)'$ symmetry-breaking may be protected by the same physics that
stabilizes the electroweak hierarchy \cite{ArkaniHamed:2008qp}.
Indeed, if the largest symmetry-breaking effects arise
from weak-scale supersymmetry breaking, then the $U(1)'$
symmetry breaking scale is naturally suppressed by a loop factor or by
$\sqrt{\epsilon}$, leading to MeV to GeV-scale $A'$ masses
\cite{ArkaniHamed:2008qp,Dienes:1996zr,Cheung:2009qd,Morrissey:2009ur,Katz:2009qq}.  

\begin{figure*}
\includegraphics[width=0.45\textwidth]{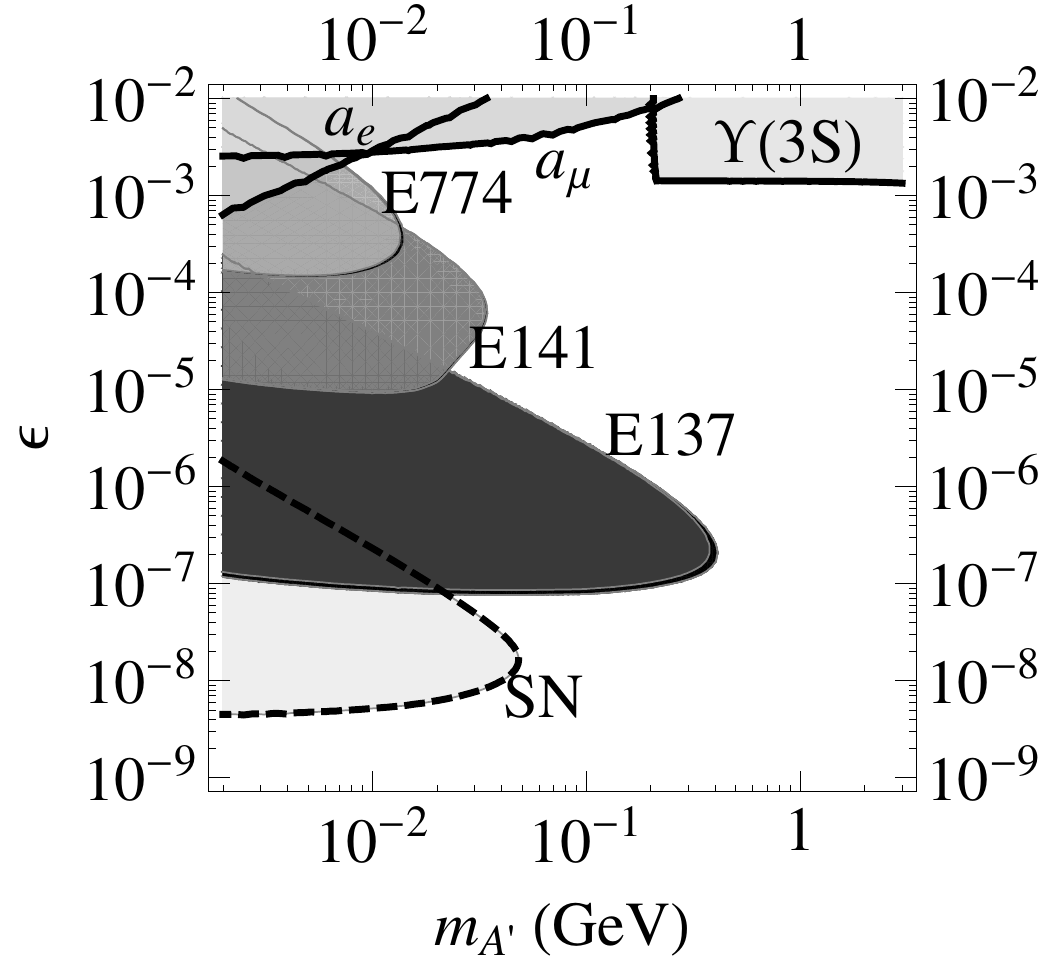}
\includegraphics[width=0.47\textwidth]{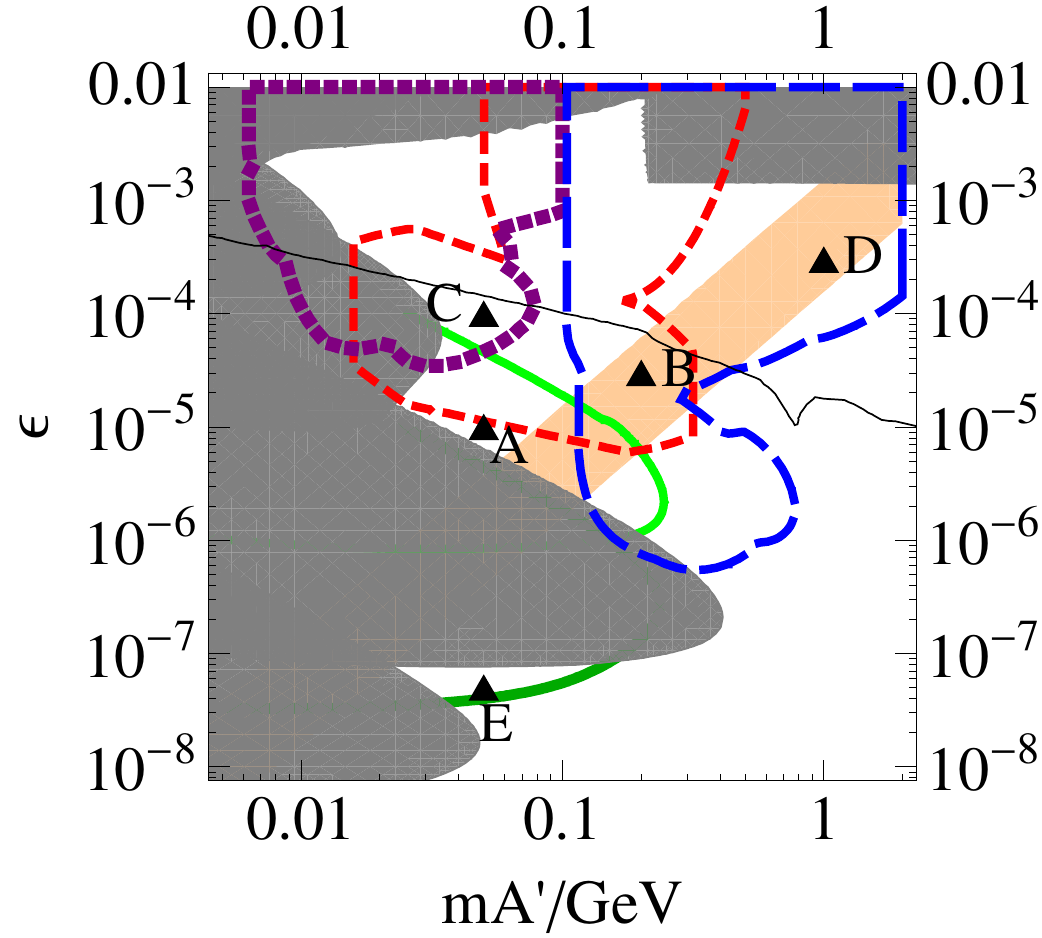}
\caption{{\bf Left:} Existing constraints on an $A'$.
Shown are constraints from electron and muon anomalous magnetic moment measurements, 
$a_e$ and $a_{\mu}$, the BaBar search for $\Upsilon(3S)\to \gamma\mu^+\mu^-$, 
three beam dump experiments, E137, E141, and E774, and supernova cooling (SN).  
These constraints are discussed further in Section \ref{sec:constraints}.  
{\bf Right:} Existing constraints are shown in gray, while 
the various lines --- light green (upper) solid, red short-dashed, purple dotted, blue 
long-dashed, and dark green (lower) solid --- 
show estimates of the regions that can be explored with the experimental scenarios 
discussed in Section \ref{subsec:expA}--\ref{subsec:expE}, respectively.
The discussion in \ref{sec:designs} focuses on the five points labeled  ``A'' through ``E''.
The orange stripe denotes the ``D-term'' region introduced in section \ref{sec:TheoryPrelim},
in which simple models of dark matter interacting with the $A'$ can explain 
the annual modulation signal reported by DAMA/LIBRA.  
Along the thin black line, the $A'$ proper lifetime $c\tau=80 \mu$m, 
which is approximately the $\tau$  proper lifetime.  
\label{fig:bigSummary}}
\end{figure*}

An $A'$ can 
be produced in collisions of charged particles with nuclei and can decay to electrons or muons.  
The production cross-section ($\sigma_{A'}$) and decay length ($\gamma c \tau$),
\bea
\sigma_{A'} &\sim& 100 \ \mbox{pb} \left (\epsilon/10^{-4} \right )^2 \left ( 100 \MeV/m_{A'} \right )^2 \\
\gamma c\tau &\sim& 1 \ \mbox{mm} \left ( \gamma/10 \right ) \left (10^{-4}/\epsilon \right )^2 \left ( 100 \MeV/m_{A'} \right )
\eea
vary by ten orders of magnitude for the $\epsilon$'s and masses $m_{A'}$ we consider. 
This wide range calls for multiple experimental approaches, with
different strategies for confronting 
backgrounds.  Beam-dump searches from the 1980's exclude the low-mass
and small-$\epsilon$ parameter range, and other data constrains large
$\epsilon$.  In this paper we suggest five scenarios for fixed-target
experiments sensitive to distinct but overlapping regions of parameter
space (see Figure \ref{fig:bigSummary}).  Together they can probe six
decades in $A'$ coupling and three decades in $A'$ mass with existing
beam energies and intensities.  

Dark matter interpretations of recent astrophysical and terrestrial
anomalies provide a further impetus to search for new $U(1)$'s.
Annihilation of dark matter charged under a new $U(1)'$ into the $A'$ 
can explain the electron
and/or positron excesses observed by PAMELA \cite{Adriani:2008zr},
ATIC \cite{2008zzr}, FERMI \cite{Abdo:2009zk}, and HESS
\cite{Collaboration:2008aaa,Aharonian:2009ah}
(see e.g.~\cite{ArkaniHamed:2008qn,Pospelov:2008jd,Hisano:2003ec,MarchRussell:2008yu,Cirelli:2008pk,Cholis:2008wq,Cholis:2008qq,Cui:2009xq}).
If the dark matter is also charged under a non-abelian group,
then its spectrum naturally implements an inelastic dark matter
scenario \cite{TuckerSmith:2001hy}, thereby explaining the annual modulation
signal reported by DAMA/LIBRA \cite{Bernabei:2005hj,Bernabei:2008yi} 
and reconciling it with the null results of other experiments 
\cite{TuckerSmith:2001hy,Chang:2008gd,ArkaniHamed:2008qn,Schuster:2009}.

%In view of the suggestive findings of these experiments, and the
In view of these suggestive data and the 
abundant theoretical speculation surrounding them, insight from new
experiments is clearly called for. New probes of weakly mixed MeV--GeV
$U(1)$'s 
directly probe the low-energy structure of these scenarios,
where the nature of their interactions is most manifest.  As such, the
experiments we advocate here are complementary to upcoming gamma-ray
observations (see e.g.~\cite{Essig:2009jx,Arvanitaki:2008hq}) and to the 
next generation of direct detection experiments 
\cite{Majorovits:2004fa,Angle:2007uj,Hasty:2006yn,Kastens:2009pa,Finkbeiner:2009ug}
that will shed light on the scattering of dark matter.

%%%%%%%%%%%%%%%%%%%%
\section*{Direct Tests of Low-Mass Gauge Sectors}\label{sec:IntroDirectTests}

Constraints on new $A'$s and the reach of different experiments are
summarized in Figure \ref{fig:bigSummary}.  To begin, low-energy
$\epm$ colliders are a powerful laboratory for the study of an $A'$
with $\epsilon \gtrsim 10^{-4}$ and mass above $\sim 200$ MeV,
particularly in sectors with multiple light states
\cite{Essig:2009nc,Batell:2009yf, Reece:2009un,Bossi:2009uw,Yin:2009mc}.  Their reach in
$\epsilon$ is limited by luminosity and irreducible backgrounds.
However, an $A'$ can also be produced through bremsstrahlung off an
electron beam incident on a fixed target \cite{Reece:2009un}.  This
approach has several virtues over colliding-beam searches: much larger
luminosities, of $\mathcal{O}(1 \,{\rm ab}^{-1}/{\rm day})$ can be
achieved, scattering cross-sections are enhanced by nuclear charge
coherence, and the resulting boosted final states can be observed with
compact special-purpose detectors.

Past electron ``beam-dump'' experiments, in which a detector looks for
decay products of rare penetrating particles behind a stopped electron
beam, constrain $\gtrsim 10$ cm vertex displacements and $\epsilon
\gtrsim 10^{-7}$.  The thick shield needed to stop beam products
limits these experiments to long decay lengths, so thinner targets are
needed to probe shorter displacements (larger $\epsilon$ and
$m_{A'}$).  However, beam products easily escape thin targets and
constitute a challenging background in downstream detectors.

The five benchmark points labeled ``A'' through ``E'' in Figure
\ref{fig:bigSummary} (right) require different approaches to these
challenges, discussed in Section \ref{sec:designs}.  We have estimated
the reach of each scenario, summarized in Figure \ref{fig:bigSummary}
(right), in the context of electron beams with 1--6 GeV energies,
nA--$\mu$A average beam currents, and run times $\sim 10^6$ s. Such
beams can be found for example at the Thomas Jefferson National
Accelerator Facility (JLab), the SLAC National Accelerator Laboratory,
the electron accelerator ELSA, and the Mainzer Mikrotron (MAMI).

The scenarios for points A and E use $100$ MeV--1 GeV electron beam dumps, 
with more complete event reconstruction or higher-current
beams than previous dump experiments.  Low-mass,
high-$\epsilon$ regions (e.g. B and C) produce boosted $A'$ and
forward decay products with mm--cm displaced vertices.  Our approaches
exploit very forward silicon-strip tracking to identify these
vertices, while maintaining reasonable occupancy --- a limiting
factor.  At still higher $\epsilon$, no displaced vertices are
resolvable and one must take full advantage of the kinematic
properties of the signal and background processes, including the
recoiling electron, using either the forward geometries of B and C or
a wider-angle spectrometer (e.g. for point D).  Spectrometers operating
at various laboratories appear capable of probing this final region.

We focus on the case where the $A'$
decays directly to Standard Model fermions, but the past experiments
and proposed scenarios are also sensitive (with different exclusions)
if the $A'$ decays to lighter $U(1)'$-charged scalars, and to direct
production of axion-like states.
%%%%%%%%%%%%%%%%%%%%%%%%%
\vspace{-0.2cm}
\section*{Outline}
\vspace{-0.2cm}
In Section \ref{sec:production}, we summarize the properties of 
$A'$ production through bremsstrahlung in fixed-target collisions.  
Constraints from past experiments and from neutrino emission by SN 1987A 
are presented in Section \ref{sec:constraints}.  In Section
\ref{sec:designs}, we describe the five new experimental scenarios and
estimate the limiting backgrounds. We conclude in Section
\ref{sec:discussion} with a summary of the prospects for new
experiments.
More detailed formulas, which we use to calculate 
our expected search reaches, and a more detailed discussion of some 
of the backgrounds, are given in Appendices \ref{app:Details}, \ref{app:moreproduction}, 
and \ref{app:backgrounds} .  

%%%%%%%%%%%%%%%%%%%%%%%%%%%%%%%%%%
%%%%%%%%%%%%%%%%%%%%%%%%%%%%%%%%%%
\section{The Physics of New $U(1)$ Vectors in Fixed Target Collisions}\label{sec:production}

%%%%%%%%%%%%%%%%%%%%
\subsection{Theoretical Preliminaries}\label{sec:TheoryPrelim}

Consider the Lagrangian 
\be\label{eq:Lag} \mathcal{L} =
\mathcal{L}_{\textrm{{SM}}} + \epsilon_Y F^{Y,\mu\nu} F'_{\mu\nu} +
\f{1}{4} F'^{,\mu\nu}F'_{\mu\nu} + m_{A'}^2 A'^{\mu} A'_{\mu}, 
\ee
where $\mathcal{L}_{\textrm{{SM}}}$ is the Standard Model Lagrangian,
$F'_{\mu\nu} =\partial_{[\mu}A'_{\nu]}$, and $A'$ is the gauge field
of a massive dark $U(1)'$ gauge group \cite{Holdom:1985ag}.  
The second term in (\ref{eq:Lag}) is the kinetic mixing operator, 
and $\epsilon\sim 10^{-8}-10^{-2}$ is naturally generated by loops 
at any mass scale of heavy fields charged under both $U(1)'$ and
$U(1)_Y$; the lower end of this range is obtained if one or both $U(1)$'s are 
contained in grand-unified (GUT) groups, since then $\epsilon$ 
is only generated by two-or three-loop GUT-breaking effects.  

A simple way of analyzing the low-energy effects of the $A'$ is to treat 
kinetic mixing as an insertion of $p^2 g_{\mu\nu} -p_\mu p_\nu$ in
Feynman diagrams, making it clear that the $A'$ couples to the 
electromagnetic current of the Standard Model through the photon. 
This picture also clarifies, for example, that new interactions
induced by kinetic mixing must involve a massive $A'$ propagator,
and that effects of mixing with the $Z$-boson are further suppressed by
$1/m_Z^2$.  Equivalently, one can redefine the photon field $A^\mu
\rightarrow A^\mu + \epsilon A'^\mu$ as in \cite{Baumgart:2009tn},
which removes the kinetic mixing term and generates a coupling
$e A_{\mu} J^{\mu}_{\rm EM} \supset \epsilon e A'_\mu J^\mu_{\rm EM}$
of the new gauge boson to electrically charged particles (here
$\epsilon \equiv \epsilon_Y \cos\theta_W$).  
Note that this does not induce
electromagnetic millicharges for particles charged under the $A'$.
The parameters of concern in this paper are $\epsilon$ and $m_{A'}$.

We now explain the orange stripe in Figure \ref{fig:bigSummary} 
--- see \cite{Dienes:1996zr,Cheung:2009qd,Morrissey:2009ur} for more details.  
In a supersymmetric theory, the kinetic mixing operator induces a mixing between 
the D-terms associated with $U(1)'$ and $U(1)_Y$.
The hypercharge D-term gets a vacuum expectation value from electroweak symmetry 
breaking and induces a weak-scale effective Fayet-Iliopoulos term for $U(1)'$.
Consequently, the Standard Model vacuum can break the 
$U(1)'$ in the presence of light $U(1)'$-charged degrees of freedom,
giving the $A'$ a mass,
\be\label{eq:DtermLine}
m_{A'} \sim \sqrt{\epsilon g_D}\, \f{\sqrt{g_Y}m_W}{g_2},
\ee 
where $g_D$, $g_Y$, and $g_2$ are the the $U(1)'$, $U(1)_Y$, and Standard Model 
$SU(2)_L$ gauge couplings, respectively, and $m_W$ is the W-boson mass.   
Equation (\ref{eq:DtermLine}) relates $\epsilon$ and 
$m_{A'}$ as indicated by the orange stripe in Figure \ref{fig:bigSummary} for $g_D\sim 0.1-1$. 
This region is not only theoretically appealing, but also 
roughly corresponds to the region in which the annual modulation signal observed by 
DAMA/LIBRA can be explained by dark matter, charged under the $U(1)'$, scattering inelastically 
off nuclei through $A'$ exchange. We therefore include these lines for reference in our plots. 

%%%%%%%%%%%%%%%%%%%%
\subsection{$A'$ Production in Fixed-Target Collisions}\label{sec:Production}

\begin{figure}
\includegraphics[width=0.275\textwidth]{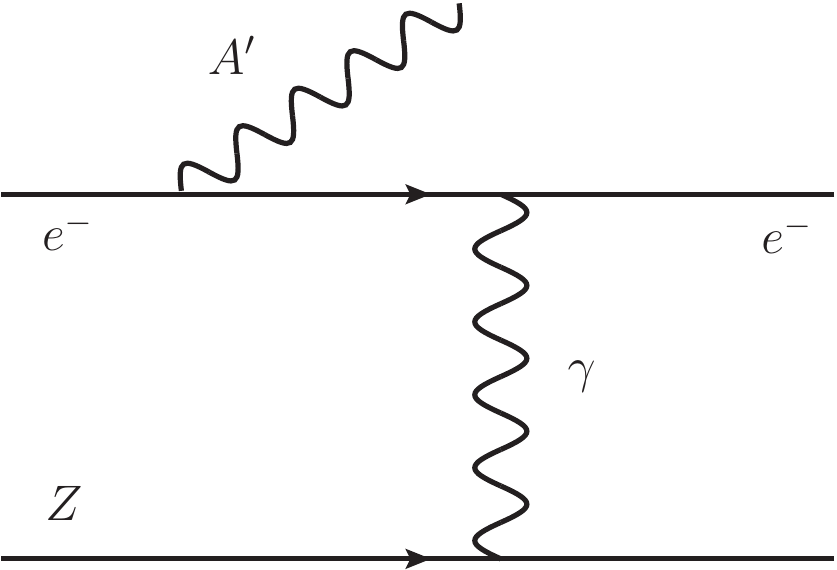}
\caption{$A'$ production by bremsstrahlung off an incoming electron 
scattering off protons in a target with atomic number $Z$.
\label{fig:Sig}}
\end{figure}

\begin{figure}
\includegraphics[width=0.45\textwidth]{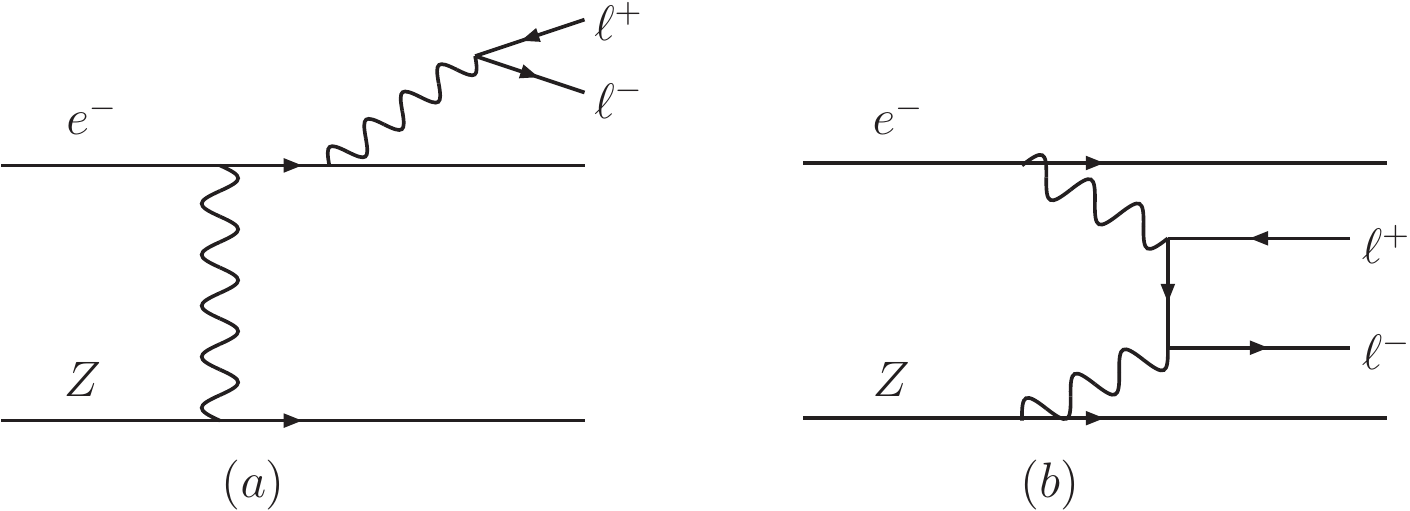}
\caption{(a) $\gamma^*$ and (b) Bethe-Heitler 
trident reactions that comprise the primary QED background to $A'\rightarrow \ell^+\ell^-$ search channels. 
\label{fig:SigAndBkg}}
\end{figure}

$A'$ particles are generated in electron collisions on a fixed target
by a process analogous to ordinary photon bremsstrahlung, see Figure \ref{fig:Sig}. 
This can be reliably estimated in the Weizs\"acker-Williams approximation (see
Appendix \ref{app:Details} for more details) \cite{Kim:1973he,Tsai:1973py,Tsai:1986tx}.  
When the incoming electron has energy $E_0$, the
differential cross-section to produce an $A'$ of mass $m_{A'}$ with
energy $E_{A'} \equiv x E_0$ is
\bea
\f{d\sigma}{dxd\cos\theta_{A'}}& \approx& \f{8 Z^2\alpha^3 \epsilon^2 E_0^2
  x}{U^2} {\mathcal{L}og} \nonumber \\
&&\hspace{-0.6in} \times \bigg[  (1-x+\f{x^2}{2}) 
 - \f{x (1-x) m_{A'}^2 \left(E_0^2 x\, \theta_{A'}^2\right)}{U^2}
\bigg]\label{eq:dSigmadxdcos}
\eea
where $Z$ is the atomic number of the target atoms, $\alpha \simeq 1/137$, $\theta_{A'}$ 
is the angle in the lab frame between the emitted $A'$ and the incoming electron, the ${\cal L}og$
($\sim 5-10$ for $m_{A'} \lesssim 500$ MeV) depends on kinematics,
atomic screening, and nuclear size effects (see Appendix \ref{app:Details}
and Figure \ref{fig:chiLOG} therein), and
\be
U(x,\theta_{A'}) = E_0^2 x \theta_{A'}^2 + m_{A'}^2 \f{1-x}{x} + m_e^2 x \label{Udef}
\ee
is the virtuality of the intermediate electron in initial-state
bremsstrahlung.  The above results are valid for 
\be
m_e \ll m_{A'} \ll E_0, \qquad  x \,\theta_{A'}^2 \ll 1.
\ee

Dropping $m_e$ and performing the angular integral, we find
\be
\f{d\sigma}{dx} \approx \f{8 Z^2\alpha^3 \epsilon^2 x}{m_{A'}^2} 
\left(1 + \f{x^2}{3 (1-x)}\right) {\cal L}og \label{dSigmadX}
\ee

The $x$-scaling and singularity structure $\f{\alpha^3}{m_e^2 x}$ of
massless bremsstrahlung \cite{Tsai:1986tx} is recovered from \eqref{eq:dSigmadxdcos} with
$m_{A'}^2=0$ (the polynomial factor differs because of finite $m_{A'}$
corrections to the matrix element), but differs from the massive $A'$-strahlung in
several important ways.  We emphasize that these properties are not
particular to any matrix element, but rather is a \emph{kinematic} property common to all 
heavy-particle emission:
\begin{description}
\item[Rate:] For most $x$, $U(x,0) \sim m_{A'}^2$, so that the total $A'$
  production rate is controlled by $\f{\alpha^3 \epsilon^2}{m_{A'}^2}$.
  Therefore, it is suppressed relative to photon bremsstrahlung by
  $\sim \epsilon^2 \f{m_e^2}{m_{A'}^2}$.
\item[Energy:] $A'$ bremsstrahlung is sharply peaked at $x \approx 1$,
  where $U(x,0)$ is minimized.  When an $A'$ is produced, it carries
  nearly the entire beam energy --- in fact the median value of
  $(1-x)$ is $\sim \max\left(\f{m_e}{m_{A'}},\f{m_{A'}}{E_0}\right)$. 
\item[Angle:] $A'$ emission is dominated at angles $\theta_{A'}$ such that
  $U(x,\theta_{A'}) \lesssim 2 \,U(x,0)$ (beyond this point, wide-angle
  emission falls as $1/\theta_{A'}^4$).  For $x$ near its median value, the
  cutoff emission angle is
\be
\theta_{\rm A'\, max} \sim \max\left(\f{\sqrt{m_{A'} m_e}}{E_0},\f{m_{A'}^{3/2}}{E_0^{3/2}}\right)
\ee
which is parametrically smaller than the opening angle of the 
$A'$ decay products, $\sim m_{A'}/E_0$.  The approximation of
collinear emission is justified in many calculations.
\end{description} 

Because these features apply to bremsstrahlung of \emph{any} massive
boson, there is a simple approximate equivalence between constraints on the
$A'$ and constraints on an axion with mass $m_a = m_{A'}$ and decay
constant $f_a \sim 1.7 \times 10^{-3} \,{\rm GeV}\,C_e/\epsilon$, at least when the
constraints come from coupling to electrons.  Here the coupling constant of axions to 
electrons is assumed to be $C_e m_e/f_a$, where $C_e$ is a model-dependent coefficient.

The total number of $A'$ produced when $N_e$ electrons of initial energy $E_0$ scatter in a
target of $T$ radiation lengths is
\bea
\f{dN}{dx} = N_e \f{N_0 X_0}{A} && \int_{E_{A'}}^{E_0}
\f{dE_1}{E_1} \int_0^T dt\, I(E_1; E_0, t) \nonumber \\
&& \times\, E_0\, \f{d\sigma}{dx'}\Big|_{x'=E_{A'}/E_1},
\eea
where $x'\equiv E_{A'}/E_1$, $X_0$ is the radiation length of the target, 
$N_0 \simeq 6 \times 10^{23} \,{\rm mole}^{-1}$ is 
Avogadro's number, $A$ is the target atomic mass in g/mole, and $I$ is the energy distribution of 
electrons after passing through $t$ radiation lengths.  

After the $A'$ is produced in the target, it will travel for some distance before 
decaying back into Standard Model particles (we will assume throughout this paper that 
no other decay channels into particles charged under the $U(1)'$ are available). 
The proper lifetime of the $A'$ is
\bea\label{properLifeTime}
c \tau &&= \f{1}{\Gamma} \simeq \f{3}{N_{\rm eff} m_{A'} \alpha \epsilon^2} \nonumber \\
&&\simeq \f{80 \,\mu \mbox{m}}{N_{\rm eff}} \left (\f{10^{-4}}{\epsilon} \right )^2 
\left ( \f{100 \,\mbox{MeV}}{m_{A'}} \right ),
\eea
where we have neglected phase-space corrections and 
$N_{\rm eff}$ counts the number of available decay products 
($N_{\rm eff} = 1$ for $m_{A'} \lsim 2
m_{\mu}$ when only $A'\to e^+e^-$ decays are possible, and $2+R(m_{A'})$ 
for $m_{A'}\ge2 m_{\mu}$, where 
$R$ is defined to be the energy dependent ratio 
$\f{\sigma(e^+e^- \rightarrow \mbox{ hadrons})}{\sigma(e^+e^-  \rightarrow \mu^+\mu^-)}$ 
\cite{Amsler:2008zzb}).    
$A'$ decays will thus create displaced vertices behind the target.  
While $c\tau$ determines the typical impact parameter for these displaced tracks, 
their vertex displacements are controlled by (for the typical kinematics with $x \approx 1$), 
\bea
\ell_0 &&\equiv \gamma c\tau \simeq \f{3 E_1}{N_{\rm eff} m_{A'}^2 \alpha \epsilon^2} \nonumber \\
&&\simeq \f{0.8\mbox{cm}}{N_{\rm eff}} \left ( \f{E_0}{10 \mbox{GeV}} \right ) \!\!
\left (\f{10^{-4}}{\epsilon} \right )^2 \!\!
\left ( \f{100\, \mbox{MeV}}{m_{A'}} \right )^2,
\label{gammaCTau}
\eea
where we have again neglected phase-space corrections.

%%%%%%%%%%%%%%%%%%%%%%%
\subsection{Approximate Total Rate Formulas}\label{sec:approxRate}

From equations \eqref{dSigmadX} and \eqref{gammaCTau}, we can 
obtain simple \emph{approximate} expressions for the rate of
$A'$ production in scattering off thin targets (with $T\ll 1$) and
thick ``dump'' targets ($T \gg 1$). 
These crude
approximations are only correct within about one order of magnitude,
but they are useful in quickly mapping out regions in the large
logarithmic parameter space.  In our results we use more accurate 
expressions that also include detector acceptances, as presented in 
Appendices \ref{app:Details} and \ref{app:moreproduction}.

In the thin-target limit $T \ll 1$, the beam is
not significantly degraded as it passes through the target, 
and $I(E_1,E_0,t) \approx \delta(E_1-E_0)$.  
In this case the total $A'$ production rate scales as
\be
N \sim N_e \,\f{N_0 X_0 }{A} \,T \, \f{Z^2\alpha^3 \epsilon^2}{m_{A'}^2}\,
{\cal L}og = N_e \,\mathcal{C}\, T\, \epsilon^2 \,\f{m_e^2}{m_{A'}^2},\label{rateApproxThin}
\ee
where $\mathcal{C} \approx 5$ is only logarithmically dependent on the
choice of nucleus (at least in the range of masses where the form-factor is
only slowly varying) and on $m_{A'}$, because, roughly, $X_0\propto
\f{A}{Z^2}$ (see Appendix \ref{app:moreproduction} and
\cite{Amsler:2008zzb}).
For example, for a
Coulomb of incident electrons
\be
\f{N}{\mbox{C}} \sim 10^{6} \left ( \f{T}{0.1} \right ) \left (\f{\epsilon}{10^{-4}} \right)^2 
\left( \f{100\, \mbox{MeV}}{m_{A'}} \right)^2.
\ee
For a thick target ($T \gg 1$), production is dominated near the front
of the target and
\be
N \sim N_e\, \mathcal{C}'\, \epsilon^2\, \f{m_e^2}{m_{A'}^2},\label{rateApproxThick}
\ee
with $\mathcal{C}' \approx 10$.  When the typical
lifetime $\ell_0$ exceeds the length $L$ to the detector, a fraction
$\sim L/\ell_0$ decay before the detector, and the number of $A'$ observed is
independent of $m_{A'}$:
\be
N_{\rm obs} \sim N_e \mathcal{C}' \epsilon^2 \f{m_e^2}{m_{A'}^2} \f{L}{\ell_0} 
\sim 
N_e \mathcal{C}' \alpha \epsilon^4 \f{m_e^2 L}{E_1}
\label{rateApproxLongLivedDump}
\ee
Note that multiple interactions in the target degrade the beam energy significantly and 
induce $A'$ transverse momenta $\sim 10\mbox{ mrad } (\mbox{GeV}/E_0)^2$ for $A'$ production in the first 
radiation length.  
These transverse momenta can be significant for low-energy dumps.  

For subsequent discussions, it is useful to translate the 
signal yields into rates as a function of beam and target parameters. 
A process $X$ with cross section $\sigma(X)$ occurs with a rate
\bea
\Phi(X)\sim 0.7 \mbox{ MHz} \left [ \f{T\cdot I_{beam}}{\mbox{nA}} \,
  \f{\sigma(X)}{Z^2 \mu \mbox{b}}\right ] \label{eq:bkgrate1}, 
\eea 
where $I_{beam}$ is the average current, and $T\lsim 1$ is the target thickness
in units of radiation lengths. 
For example, $m_{A'}=100 \MeV$ and $\epsilon=10^{-4}$ gives $\sigma_{A'}\sim 0.01 Z^2$ pb, 
or a rate of $\Phi_{A'}\approx 0.007  \mbox{ Hz} \left [ \f{T\cdot I_{beam}}{\mbox{nA}} \right ]$.
In contrast, Bethe-Heitler pair production (Figure
\ref{fig:SigAndBkg}) has a total cross-section $\sigma_{l^+l^-} \sim Z^2
\mu$b, or rate $\Phi_{l^+l^-}\approx 0.7 \ \mbox{MHz} \left [ \f{T\cdot I_{beam}}{\mbox{nA}} \right ]$
for a $\sim 1\GeV$ electron beam.  
Bethe-Heitler pair production is thus a potential background for 
any $A'$ search, and so our experimental scenarios will be strongly influenced 
by the need to remove them. 

In appendix \ref{app:backgrounds}, we discuss the kinematics of Bethe-Heitler production
relative to $A'$ production in some detail, and sketch out a set of selection cuts that can be used to 
suppress the otherwise prohibitively large Bethe-Heitler backgrounds 
(Figure \ref{fig:SigAndBkg}(b)). 
However, a minimal contribution to the background is obtained by
replacing the $A'$ by a $\gamma^*$ (Figure \ref{fig:SigAndBkg}(a)).
To see this, we re-insert the dilepton invariant mass $m^2$ into the
fully differential
cross-section and consider integrating over a mass window $\delta m$,
with $\Gamma \ll \delta m \ll m$.   The $A'$ and $\gamma^*$
matrix elements are related by the substitution
\be
\frac{\epsilon^2}{m^2 - (m_{A'}^2+im_{A'}\Gamma)^2} \rightarrow \frac{1}{m^2}.
\ee
All other terms in the cross-section are slowly varying in this
window.  Treating them as constant, the integration over the mass
window bounded by $m\pm \frac{\delta m}{2}$ is straightforward.
Substituting \eqref{properLifeTime}, we obtain the ratio of fully
differential cross sections for $A'$ to $\gamma^*$ production in this
mass window, which is also an upper bound on the total signal to
background,
\be
\f{d\sigma(X\rightarrow A' Y\rightarrow l^+l^- Y)}{d\sigma(X\rightarrow \gamma^* Y\rightarrow l^+l^- Y)}
=  \left ( \f{3\pi\epsilon^2}{2N_f\alpha} \right ) \left ( \f{m_{A'}}{\delta m} \right ) \label{eq:IrreducibleBackground1},
\ee
where $N_f$ is the number of available decay species for the $A'$, and
$\delta m$ is the width assigned to the $\gamma^*$ process. 
Equation (\ref{eq:IrreducibleBackground1}) summarizes the maximum achievable signal to background ratio
that any experiment can achieve in an $A'\rightarrow \ell^+\ell^-$ search
using only kinematics, with the decay vertex unresolved. 

%%%%%%%%%%%%%%%%%%%%%%%%%%%%%%%
%%%%%%%%%%%%%%%%%%%%%%%%%%%%%%%
\section{Beam Dump Constraints and Sensitivity of Current
  Experiments}\label{sec:constraints} 

In this section, we discuss existing constraints on the $\epsilon$ versus $m_{A'}$ 
parameter space, which are summarized in Figure \ref{fig:bigSummary}.  

For $m_{A'}>2m_{\mu}$, a search for $\Upsilon(3S)\to \gamma A'\to \gamma\mu^+\mu^-$ 
by the BaBar collaboration \cite{:2009cp} rules out $\epsilon\gtrsim 10^{-3}$ 
(see also \cite{Essig:2009nc}), while the electron and muon anomalous magnetic 
moments rule out the low-mass-high-$\epsilon$ region \cite{Pospelov:2008zw}. 

Strong constraints are also obtained from electron beam-dump experiments searching for MeV-mass
axions.  The strongest constraints come from the E137 \cite{Bjorken:1988as} 
and E141 \cite{Riordan:1987aw} experiments at SLAC, and the E774 \cite{Bross:1989mp} 
experiment at Fermilab: 
\begin{description}
\item[SLAC E137] dumped 30 C of electrons at 20 GeV into aluminum
  targets \cite{Bjorken:1988as}.  Beam products traveled through a 200 m hill and an
  additional 200 m of open region before hitting a $(3 \mbox{ m})^2$
  detector.  No candidate events were observed. The contour in Figure
  \ref{fig:bigSummary} represents an expected signal of 10 events (we have
  idealized the detector as a circle of radius 1.5 m).
\item[SLAC E141] dumped $2\times 10^{15}$ electrons at 9 GeV into a
  12-cm tungsten target, with a 10-cm tungsten target used for
  calibration \cite{Riordan:1987aw}.  The detector was located 35 m
  from the dump, and the analysis required observing a single decay
  product carrying over 0.5 times the beam energy with angular
  acceptance set by a 7.5-cm pipe.  Based on the background rates
  reported by the experiment, the exclusion in Figure \ref{fig:bigSummary}
  represents an expected signal of 1000 events.
  \item[Fermilab E774] dumped $0.52\times 10^{10}$ electrons at 275 GeV 
  onto two 28-radiation-length-thick (about 19.6-cm) stacks of tungsten plates \cite{Bross:1989mp}.  
  The overall target length, including veto counters behind the target, was 30-cm.  
  An electromagnetic calorimeter with an angular acceptance of about 20-cm 
  was placed 7.25 m downstream from the dump.  
  The trigger required an energy deposition of at least 27.5 GeV and 
  no signal from the veto counters.  Based on the results reported by the experiment, 
  Figure \ref{fig:bigSummary} represents an expected signal of 17 events.   
\end{description}
The approximate formulas given in Section \ref{sec:approxRate} are
sufficient to understand the shape and magnitude of the beam-dump
limits shown in Figure \ref{fig:bigSummary}: they are bounded above by
a diagonal along which many $A'$ may be produced, but all decay within
the shielding that stops the beam, and from below by a line of
diminishing rate, which is diagonal if the typical decay occurs before
the detector position, and approximately horizontal if the average
decay length $\ell_0$ is larger than the length scale $L$ of the
experiment (see equation \ref{rateApproxLongLivedDump})).  Similar
limits can be derived for alternative $A'$ decay modes, for
example if the $A'$ decays to dark-sector higgses with typical proper
lifetime that scales as $\epsilon^{-4}$ rather than $\epsilon^{-2}$.

Supernova cooling places a significant constraint on lower $\epsilon$
and lighter $m_{A'}$.  A proper accounting of supernova limits on the
$A'$ is beyond the scope of this paper, but we outline a simple
estimate based on scaling similar results for axions
\cite{Turner:1987by}.  The hot core of a collapsing supernova can cool
through production of $A'$ if they decay $\gtrsim 10$ km from the
point of production (the mean free path is typically longer than the
lifetime).  However, neutrino observations of SN1987A
confirmed an energy loss over 5--10 seconds of $1-4 \times 10^{53}$
erg.  Following \cite{Turner:1987by}, we require the energy loss in
$A'$ emission not to exceed $10^{53} \mbox{ erg}/\mbox{s}$.  We take
the $A'$ luminosity per unit energy from the core to be
\be
\f{dL}{dE_A} \sim \f{1}{T_{SN}} (6\times 10^{70} \mbox{ erg}/\mbox{s})
e^2 \epsilon^2 
\ee
for $E_A < T_{SN}=30$ MeV, which is suppressed by
$T_{SN}/m_p$ (where $m_p$ is the proton mass) 
relative to the axion rate \cite{Turner:1987by,Iwamoto:1984ir} because
the vector emission matrix element is proportional to $v^2$, whereas the axion
emission matrix element approaches a constant as $v\rightarrow 0$.
We impose an additional Boltzmann suppression $e^{-E_A/T}$ for $E_A >
T_{SN}$ and multiply by the fraction $f(E_A)= e^{-10\mbox{
    km}/\ell_0}$ that leave the supernova core.  Requiring that the
total luminosity not exceed $10^{53} \mbox{ erg}/\mbox{s}$, we can
exclude the lower-most region in Figure \ref{fig:bigSummary}.  We
emphasize that the luminosity obtained by scaling is only correct
within an order of magnitude.  An error in the cross-section would
affect the lower limit in $\epsilon$ proportionally, but the upper
limit only logarithmically.

Constraints from other experiments are all contained within the limits
from the experiments discussed above.  For example, the region
constrained by the SLAC search for milli-charged particles, which also
used an electron beam, is contained within E137
\cite{Prinz:1998ua}.  Proton beam dumps can produce $A'$ in radiation
directly from the proton or in radiation from electrons produced by
the nuclear shower.  Both processes produce $A'$ of much lower energy
than the primary proton (hard bremsstrahlung off the proton is
suppressed by the proton's finite size, and the shower electrons are
quite soft).  These softer $A'$ typically decay inside the dump.
Therefore, proton dumps such as the CHARM experiment at CERN
\cite{Bergsma:1985qz} do not exclude new regions, though they do
overlap significantly with the E137 exclusion.  Likewise, experiments
dumping proton beams for other purposes (e.g. neutrino experiments
such as MINOS and MINIBOONE) have little or no potential reach beyond
E137.

We have also considered potential limits from $A'$ production off cosmic
rays impinging on Super-K, AMANDA, and ICE-CUBE detectors, which is
dominated by bremsstrahlung off muons near ground-level, which must
only survive $\sim 1$ km to reach the detectors.  The potential
sensitivities of these experiments are contained within the E137
excluded region.  

%%%%%%%%%%%%%%%%%%%%%%%%%%%%
%%%%%%%%%%%%%%%%%%%%%%%%%%%%
\section{Scenarios for New Experiments}\label{sec:designs}

The parameter space that new experiments must cover spans a huge range. The $A'$
production cross section and decay width vary as $\epsilon^2$ and thus
vary over ten orders of magnitude.  Our purpose in this section is to
explore experimental scenarios appropriate to different parameter
ranges.  For the sake of definiteness, we organize the discussion
around parameter points labeled ``A'' through ``E'' in Figure
\ref{fig:bigSummary}. Each choice suggests a different experimental
approach, described in the appropriately labeled subsections.

We do not intend here to provide detailed designs; this is
the task of those who would actually do the experiments.  We do
attempt to show that the exploration of this parameter space is
experimentally feasible and offer some guidance regarding how to
choose design parameters in order to optimize the experimental
sensitivity.  Because the electron beams at Jefferson
Laboratory appear to be an attractive choice for such experiments, we
have been guided in our considerations by the beam specifications
available there. However, we expect that other attractive options
exist elsewhere.

There is a natural dividing line in the parameter space
(cf.~Figure \ref{fig:bigSummary}), corresponding to an $A'$ proper 
lifetime  $c\tau\simeq 80\,\mu$m, comparable to that of the $\tau$ lepton.  
Longer lifetimes allow in principle
the determination of a separated decay vertex, while much shorter
lifetimes do not. Since determination of the detached vertex is a
strong experimental signature, the experimental techniques naturally
differ in the two regimes.  Beam-dump searches, including our first (and last)
scenario, are appropriate to much longer lifetimes.  
The region near the dividing line has not yet been explored, and
micro-vertex detectors appear quite promising in this range. The
second and third scenarios  we describe assume this
technique.

For very short lifetimes of the $A'$, the experimental signature is
identical to electromagnetic trident production $e+Z \rightarrow 3\,e$ or
$e+Z\rightarrow e+2\mu$ (where $Z$ is the target), shown in 
Figure \ref{fig:SigAndBkg}.  The 
simple upper bound discussed in equation (\ref{eq:IrreducibleBackground1})
in Section \ref{sec:production} on the ratio of the $A'$ fully differential
cross-section to the background trident rate implies that high
statistics and resolution are required to have any chance of
observing the $A'$.  Fortunately, such a regime does appear to overlap
with the capabilities of Jefferson Laboratory spectrometers.
Moreover, the upper bound is attained in a sizeable region of the
differential phase space where $A'$ production is dominant, if appropriate 
kinematic cuts are applied on the final state leptons 
(see Appendix \ref{app:backgrounds}). Therefore this parameter
region seems in principle to be accessible. Our fourth experimental
scenario deals with this case.  We will also show that the second and 
third scenarios we discuss have some interesting new reach in parameter space 
when used at lower luminosity as high-resolution forward spectrometers.  

For the smallest values of $\epsilon$, the primary consideration is
simply producing enough $A'$s to study experimentally, so beam-dump
experiments are the technique of choice. However, it becomes very
challenging to design beam dumps with average power exceeding 1
megawatt (MW). There appears to be a small window of opportunity
available for such a search, which would increase the reach beyond
that of E137. This comprises the fifth scenario that we discuss.

%%%%%%%%%%%%%%%%%%%%%%%%%%%%%
\subsection{Low Power, 10 cm Tungsten Beam Dump; $\epsilon = 10^{-5}$; 
$m_{A'} =  50$ MeV} \label{subsec:expA}
\begin{figure*}
\halfPage{\includegraphics[width=0.9\textwidth]{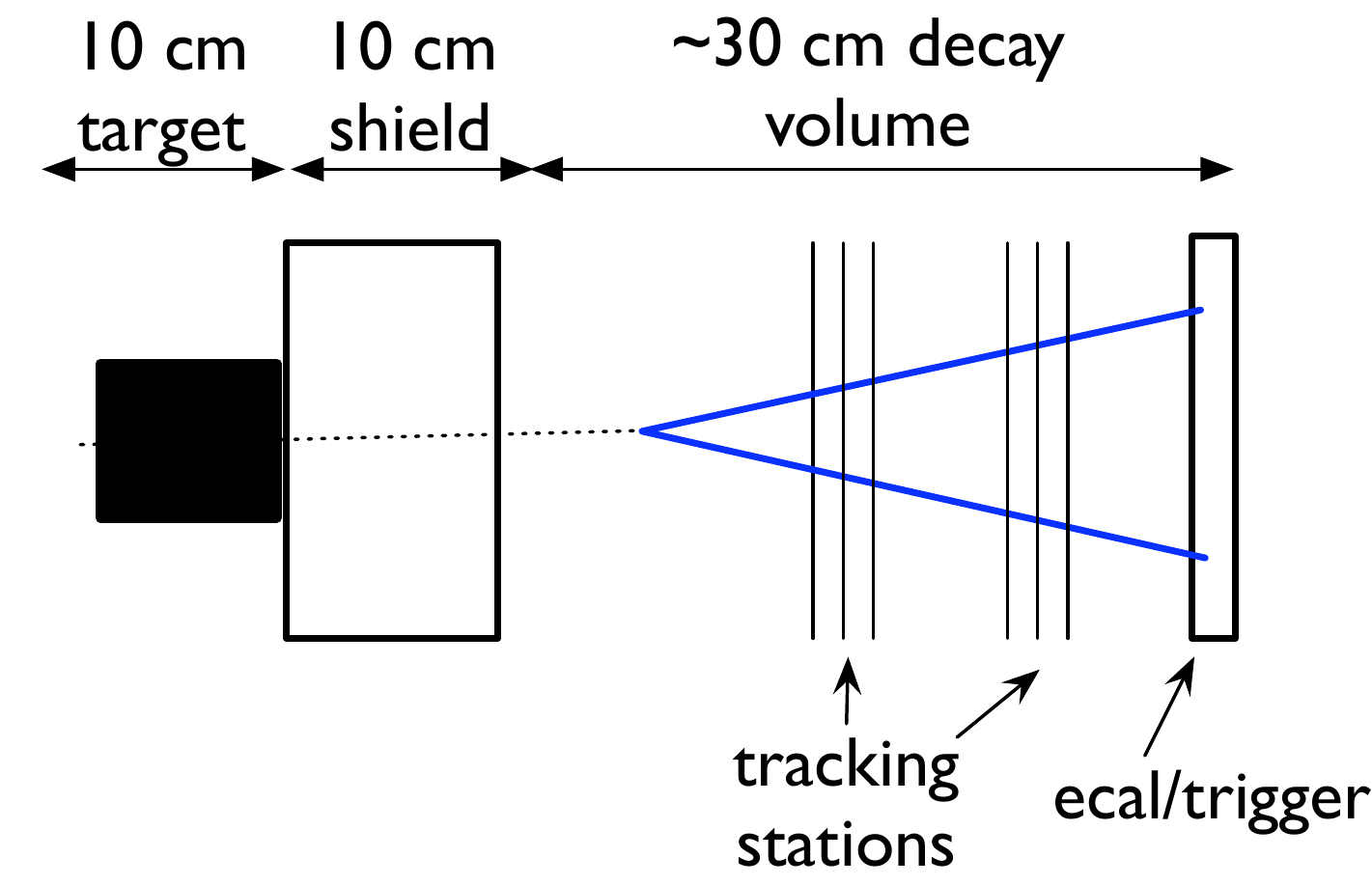}}
\halfPage{\includegraphics[width=0.9\textwidth]{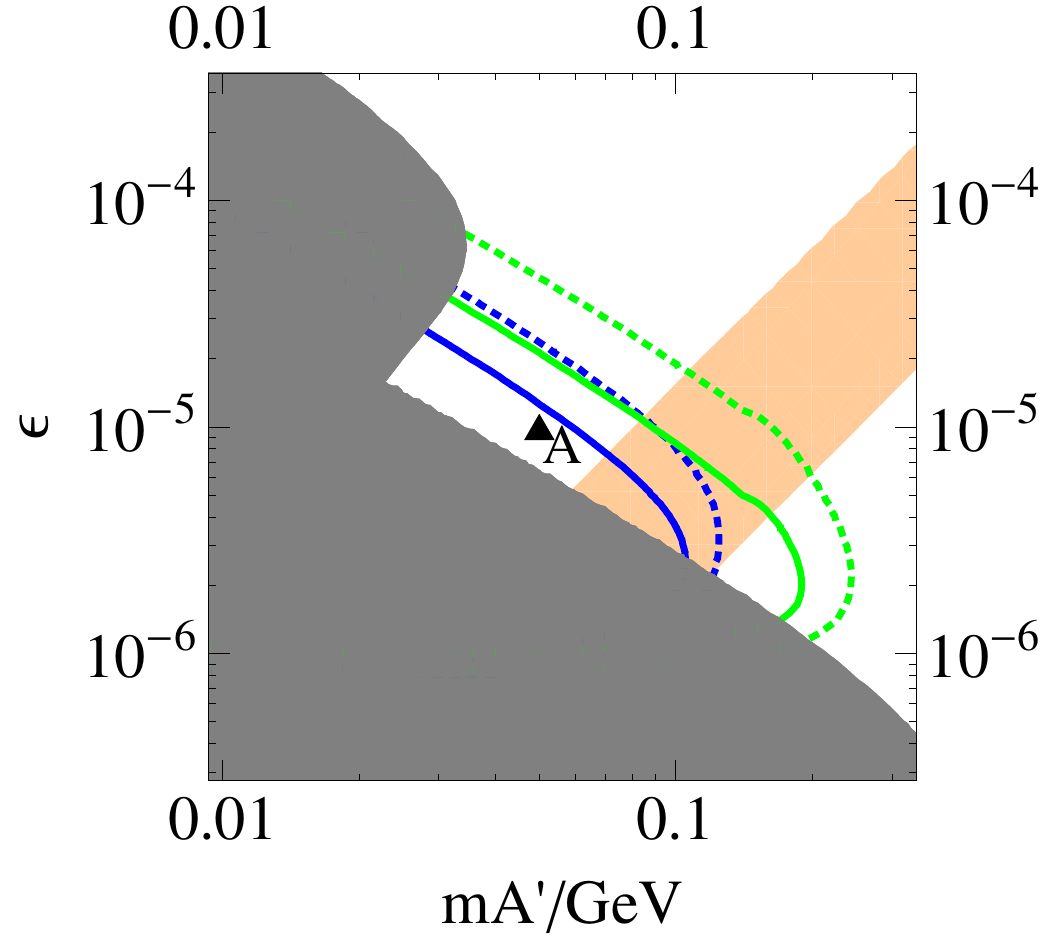}}
\caption{{\bf Left:} Experimental scenario for benchmark point A 
($\epsilon\sim 10^{-5}$, $m_{A'}\sim 50$ MeV).  
An electron beam is incident on a 10 cm thick tungsten target.  Behind the target 
is a 10 cm (or thicker) shield followed by an instrumented decay region consisting of 
a combination of tracking planes, electromagnetic calorimetry and scintillator 
triggers. 
{\bf Right:} Reaches of the high- and low-energy dump configurations
described in Section \ref{subsec:expA}, delineated by regions with 10 or more events
and the following configurations --- 
\emph{Blue (inner) Solid Contour:} $0.3$ C total charge dumped with a 200 MeV electron beam,
a 20 cm shield, and a detector with 5 cm radius 50 cm behind the front of the target.  
The lepton pair must have total energy exceeding 100 MeV.
\emph{Blue (inner) Dashed Contour:} same configuration, but with no shield.  
\emph{Green (outer) Solid Contour:} $0.1$ C (100 nA beam $\times$ $10^6$ s) 
total charge dumped with a 6 GeV electron beam, a 3.9 m shield, and a detector with 10 cm radius 7 m downstream.  
The lepton pair must have total energy exceeding 3 GeV.  
\emph{Green (outer) Dashed Contour:} same configuration, but with 0.9 m of shielding.   
 \emph{Gray contours and Orange Stripe:} exclusions from past experiments (E137 and E141)
 and the region that explains DAMA/LIBRA in a simple model  --- 
see Figure \ref{fig:bigSummary} for more details.
\label{fig:LPDumpReach}}
\end{figure*} 
  
We consider a 200 MeV primary electron beam incident on a
10 cm tungsten target.  Downstream (beyond a thin, but dense shielding
wall, if necessary), is an instrumented decay volume containing a
combination of tracking planes, electromagnetic calorimetry, and
scintillator triggers. With the chosen values of $\epsilon$ and $m_{A'}$, the
laboratory decay length of a typical $A'$ of momentum 160 MeV is about
5 cm. The produced $A'$s are contained in an angular
cone of order 125 mrad. Therefore the tracking
system in an experimental region no more than 40 cm (see Figure \ref{fig:LPDumpReach})
downstream of the front of the dump need have transverse dimensions no
more than 10 cm to identify the $A'$ decay vertices and measure the angles. Since the
decay angles of the electron/positron pair are of order 250 mrad, the
calorimeter transverse dimensions can be very modest.

In this scenario, the total yield per incident electron of $A'$s
containing at least 80 percent of the beam momentum is about $9\times
10^{-15}$ per electron dumped.  If the front of the fiducial decay
volume can be located immediately behind the 10 cm target, $5\%$ of
these $A'$s decay outside the target for a rate of $4\times 10^{-16}$
per electron dumped.  If a thicker shield is necessary to stop soft
photons, the yield remains large: $0.1\%$ of the $A'$ decay beyond 30
cm from the front of the dump, for one observable $A'$ decay per
$5\times 10^{16}$ electrons dumped.  In the conservative
configuration, with a total of 30 cm of material, a yield of order 30
events would be observed for 0.3 coulombs of electrons dumped (300 nA
in an experiment of duration $10^6$ seconds). This requires a modest
60 watts of beam power on the tungsten dump.

The length of the fiducial decay region need only be 20 centimeters to
capture the majority of the $A'$s emerging from the dump into the
decay region. The compact nature of this decay volume suggests the
possible use of silicon strip detectors for the tracking system.

The question of backgrounds must of course be addressed. A fast, dense
tracking system seems to be appropriate. With a readout rate at least
10 MHz, and with a continuous incident beam such as exists at
Jefferson Laboratory, there would be about 30000 electrons dumped per
readout cycle. The shower products should be absorbed efficiently, and
it should be very rare that a prompt calorimeter signal of more than
100 MeV energy deposition occurs, especially because the time
resolution of a scintillator trigger/electromagnetic calorimeter
system will be much better than 100 ns.  The residual problems,
beyond the scope of this sketch, probably have to do with neutrons and
soft photons or x-rays. 

Evidently, if we lower $m_{A'}$ and increase $\epsilon$ in such a way as
to decrease neither the rest-frame decay length nor the production
rate, the experiment will be easier, since $\gamma c\tau$ is larger and 
more $A'$ will decay in the detector volume. In Figure \ref{fig:LPDumpReach}, 
we present our rough estimate of the region of parameter space 
accessible to the experiment as described.

To extend the reach to larger masses and smaller
$\epsilon$, higher beam energies are required.  Once the threshold for
electro-production of muons and hadrons has been crossed, the
experiment may require a thicker shield, and much higher energies
appear advantageous.  For example, consider raising $m_{A'}$ and
lowering $\epsilon$ by a factor of 2, to 100 MeV and $5\times
10^{-6}$.  For a beam energy of 6 GeV, the
decay length is 2 m. If a 3 m decay volume can
be positioned with its upstream end within 4 m of the front of
the dump, then a larger fraction of the produced $A'$s can be detected
than our original example. This in turn lessens the beam-intensity
requirement; an average current of 100 nA, which leads to a dump power
of under 1 kilowatt, appears to suffice.

The detector geometry can simply be a longitudinally stretched version
of the previous case, with transverse dimensions again quite small, of
order 15 -- 20 cm. However, new backgrounds appear. Muons will
penetrate the decay volume as well as electromagnetic showers
initiated within the hadronic cascade. One leading candidate for
background troubles comes from electro-production of the $\rho$, with a
leading charged pion from the rho decay undergoing a charge-exchange
reaction into a $\pi^0$ a few radiation lengths in front of the
detector region. We have used the experience obtained in E141 to make
rough estimates, which indicate that such backgrounds are
surmountable. But the soft backgrounds such as neutrons and hard x-rays
also need to be carefully studied.  

%%%%%%%%%%%%%%%%%%%%%%%%%%%%%%%%%%%%
\subsection{Thin Target and Double Arm Spectrometer; $\epsilon = 3\times 10^{-5}$; $m_{A'} = 200$ MeV}  
\label{subsec:expB}

\begin{figure*}
\halfPage{\includegraphics[width=0.9\textwidth]{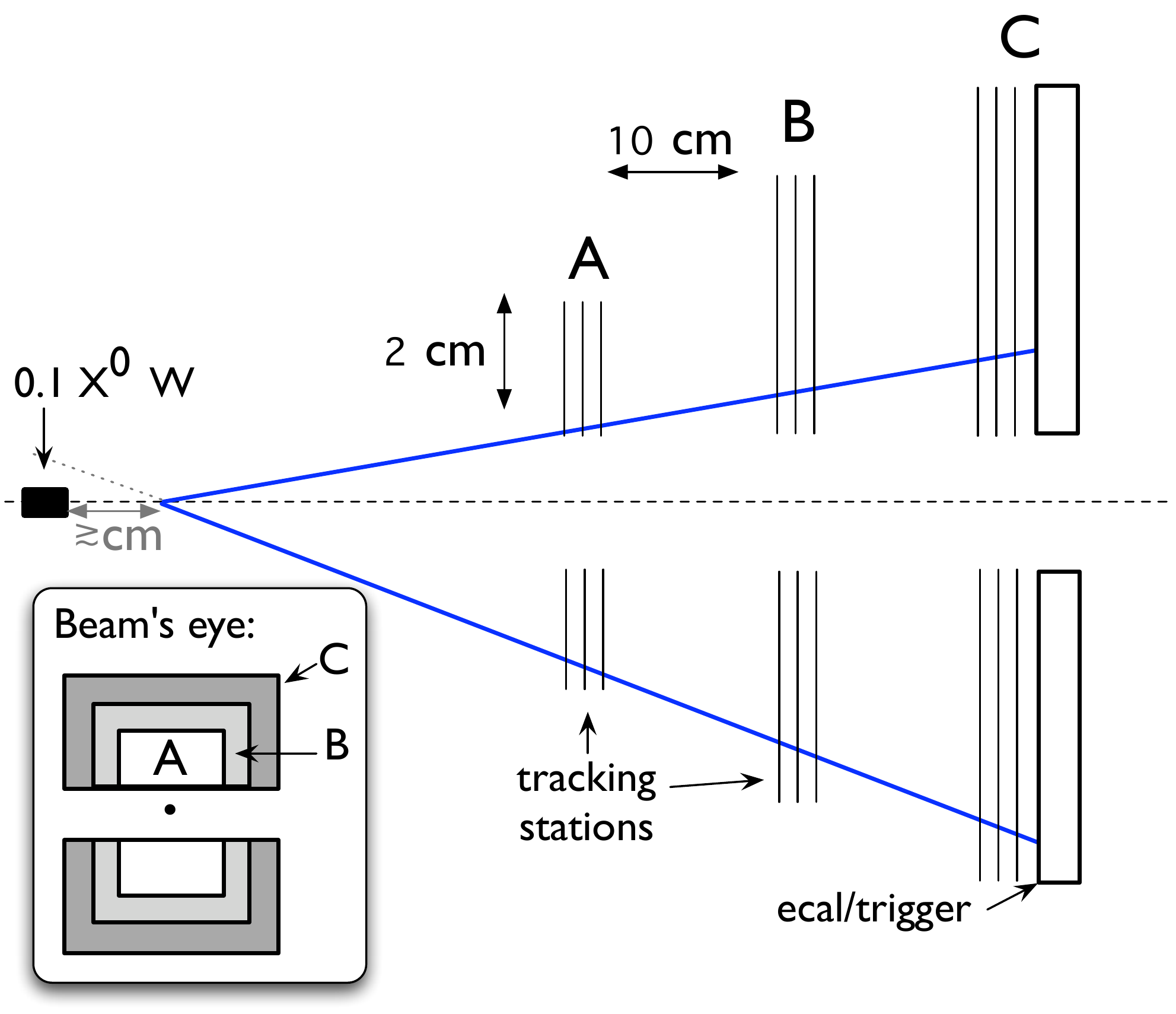}}
\halfPage{\includegraphics[width=0.9\textwidth]{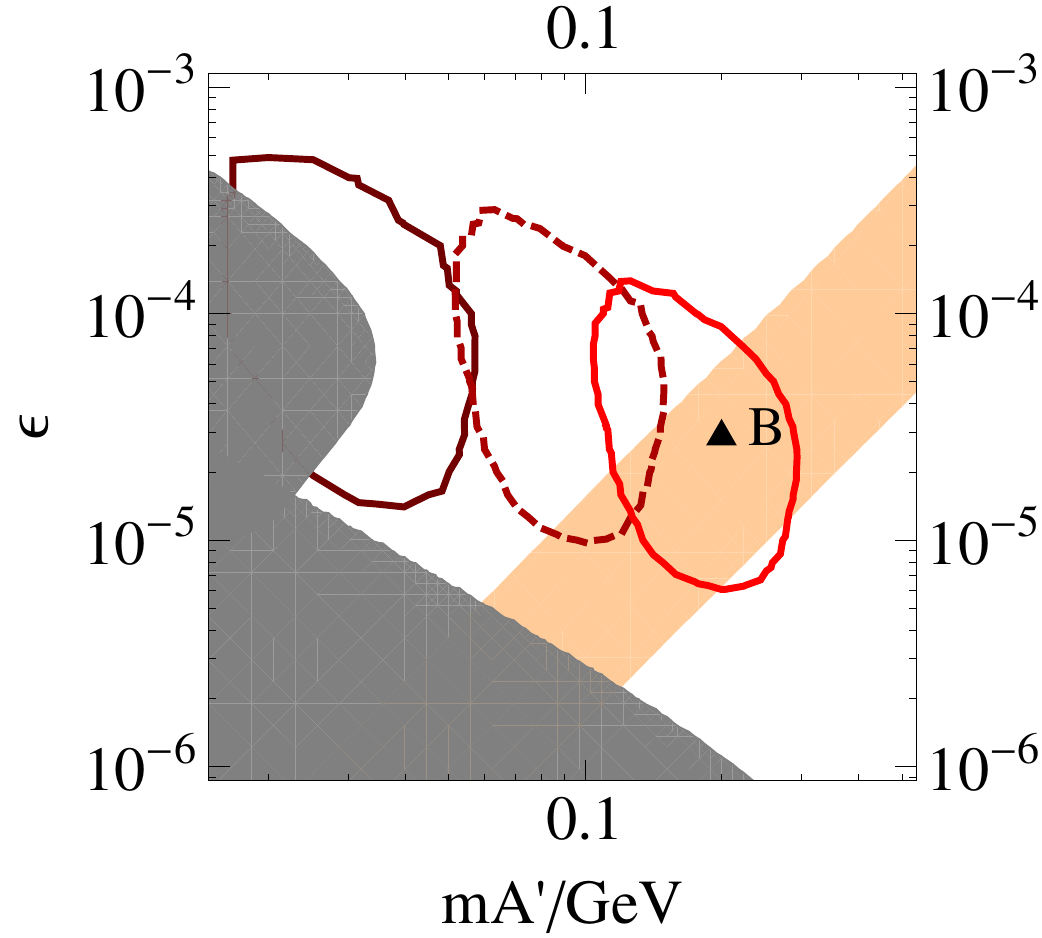}}
\caption{ {\bf Left:} Experimental scenario for a small two-arm spectrometer for benchmark point B 
($\epsilon\sim 3\times 10^{-5}$, $m_{A'}\sim 200$ MeV).  
An electron beam is incident upon a thin 0.1 radiation length tungsten target.  
A small two-arm spectrometer with silicon-strip trackers 
and a fast calorimeter or scintillator trigger is downstream from the target.  
Signal events are identified by requiring a displaced vertex $\sim$ 1 cm behind the target.  
More details are given in the text. 
{\bf Right:} 
Regions corresponding to 10 or more events within acceptance in
$10^6$ sec for three different geometries.  \emph{From right to left:} 6
GeV electron beam at 100 nA (0.1 C delivered), 
with angular acceptance from 20 to 55 mrad and a 1 m long detector (\emph{solid 
red line}); 6 GeV beam at 5 nA ($5\times 10^{-3}$ C delivered), with angular 
acceptance from 10 to 27 mrad in a 2 m-long detector region (\emph{dashed darker red line}); 
and 2 GeV beam at 0.5 nA ($5\times 10^{-4}$ C delivered) 
with the same geometry as the dashed red line (\emph{solid dark red line}).  
In all cases, we require that the $A'$ carry at least 83\% of the beam
energy, the track impact parameters at the target exceed 50 $\mu$m,
and the reconstructed vertex displacement exceed 1 cm.  We 
assume 50\% $\phi$ coverage.
\emph{Gray contours and Orange Stripe:} exclusions from past experiments (E137 and E141)
and the region that explains DAMA/LIBRA in a simple model  --- 
see Figure \ref{fig:bigSummary} for more details.
\label{fig:ForwSpec}}
\end{figure*}

Modern micro-vertex detectors allow much better lifetime resolution
than the above example. When $\epsilon$ is increased from the previous
example, the rate of $A'$ production per incident electron increases, 
and a thin target can be
used instead of a beam dump. For the parameters of interest here, we
consider a 0.1 radiation length tungsten target. We choose a 6 GeV
beam with an average current of 100 nA. Downstream of the target
is a two-arm mini-spectrometer with silicon strip detectors as the
tracking elements, backed up with fast calorimeter/scintillator
triggers. 

With these parameters, the $A'$ production rate (before acceptance)
out of the target is about 10 per hour. The angular divergence of the
$A'$ beam is only about 5 mrad. The laboratory decay length is about
1 cm, and the decay products of the $A'$ have an average angle of
about 35 mrad from the beam axis. A spectrometer with polar angle
coverage of 20 to 55 mrad and 50\% azimuthal angle coverage has about
25\% acceptance for the $A'$ decay products. The trigger
requirement includes the demand that the energies in each of the
calorimeters are between 1 and 5 GeV, with the sum between 5 and 6
GeV. The tracking system must identify one track in each arm that
points to the calorimeter hit (if the calorimeter is segmented) and
is consistent with a decay-vertex origin. After reconstruction,
additional kinematic constraints provide rejection
power.  In Figure \ref{fig:ForwSpec}, we show the reach of this 
experimental scenario for various geometries and different beam currents.   

A major background is simultaneous elastic coulomb scattering in each
arm.  An elastically scattered electron deposits 6 GeV in the
calorimeter, and is rejected, but the singles rate must be below one
per timing window (100 MHz or less for fast calorimeters).  This
requirement is safely met by the beam intensity quoted above. The
elastic-scattering radiative tails will contribute to the trigger, but
at a significantly lower rate of 10 kHz or so. Other sources for
background triggers, such as Bethe-Heitler pair production (cf.~Figure
\ref{fig:SigAndBkg}), lead to smaller or comparable trigger rates.
When one of the two scattered electrons scatters again in the first
layer of silicon, the intersection of the two reconstructed tracks is
displaced.  We find that the rate for these fake vertices is
adequately suppressed if the first layer is placed close to the
target, within $\sim 5-10$ cm.

Another basic requirement is that the occupancy in the tracking system
be acceptably low. High-resolution silicon strip detectors are
beneficial in this regard. Within a cone of opening angle of 10 mrad
at a distance of 50 cm downstream of the target, we estimate that the density of 
electrons and photons produced in the target with energy above 1 MeV is of order
$10^9/\mbox{cm}^2/\mbox{s}$
\footnote{We thank Takashi Maruyama for providing calculations of
  downstream particle densities using FLUKA.}.  In this scenario, the
silicon is placed further from the beam, but this rate serves as a
rough upper bound, which would give one percent occupancy for a 1 cm
$\times$ 25 $\mu$m strip. While these numbers are encouraging, a
serious simulation is certainly required.

%%%%%%%%%%%%%%%%%%%%%%%%%%%%%%%
\subsection{Silicon Strip Layers in a Diffuse Electron Beam; $\epsilon
  =10^{-4}$; $m_{A'} = 50$ MeV } \label{subsec:expC}

\begin{figure*}
\includegraphics[width=0.45\textwidth]{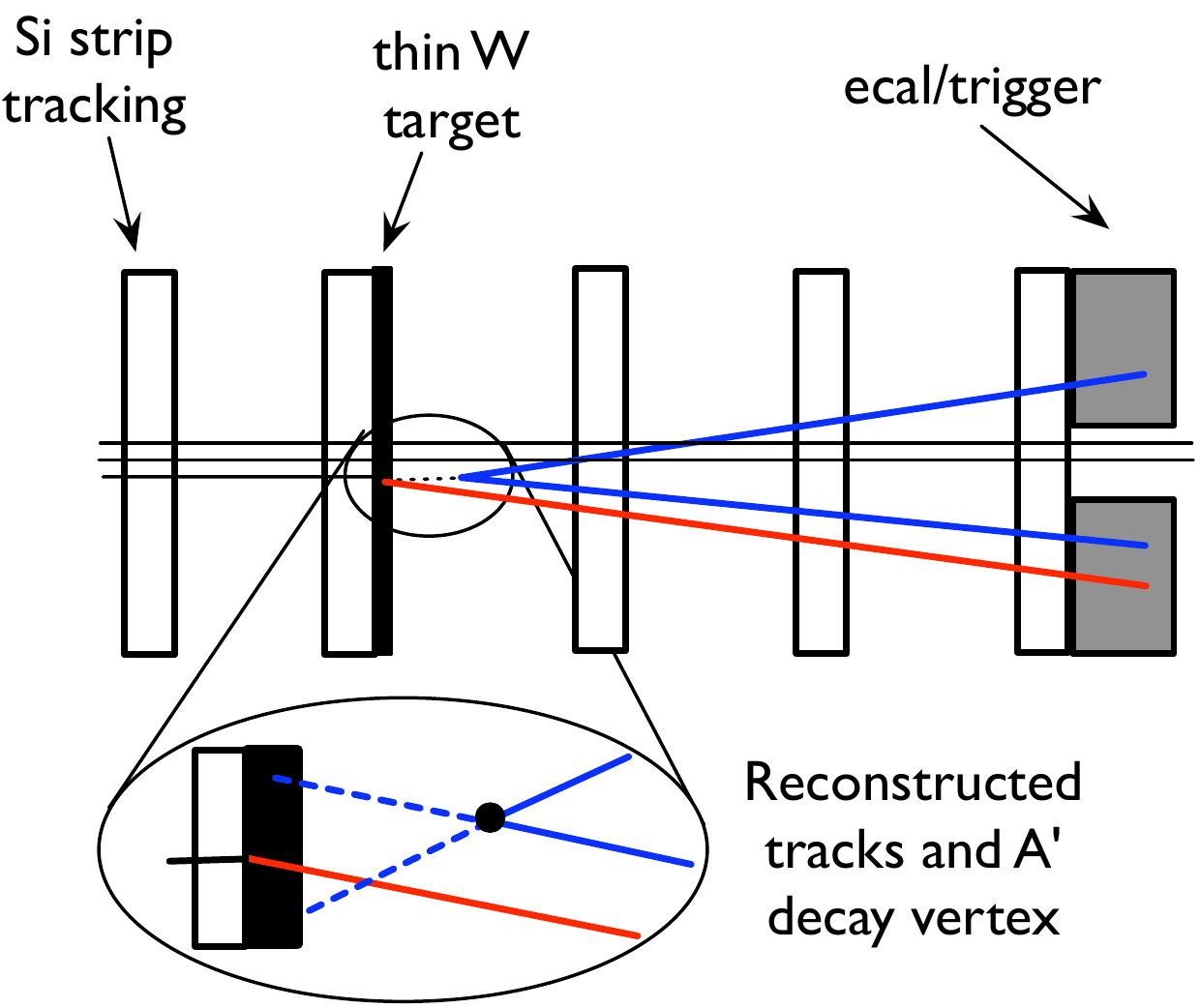}
\includegraphics[width=0.45\textwidth]{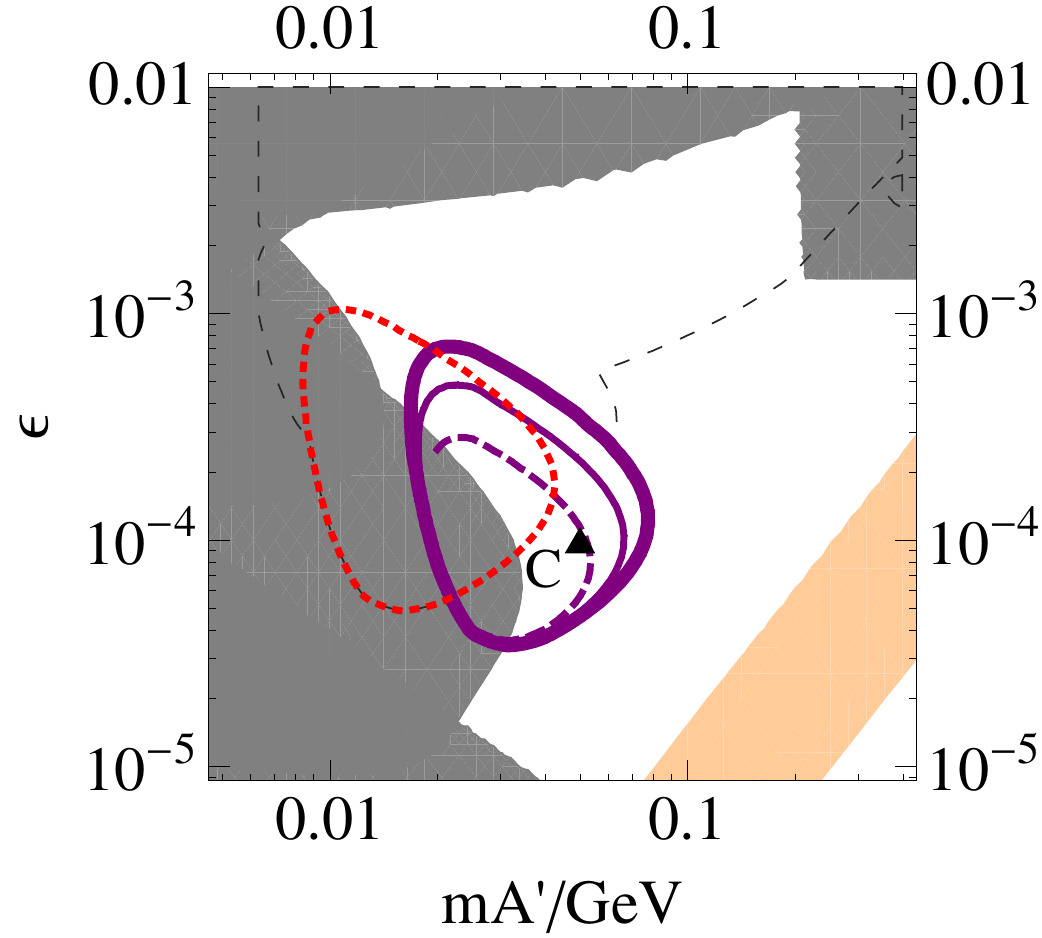}
\caption{{\bf Left:} Experimental scenario for benchmark point C 
($\epsilon\sim 10^{-4}$, $m_{A'}\sim 50$ MeV). 
Silicon strip tracking elements, together with a 0.1 radiation length ($300 \mu$m) tungsten 
target directly behind one of the elements, are inserted into a 1 GeV diffuse 
(1 cm $\times$ 1 cm) electron beam of intensity $\lsim 10^8$ e$^-$/s.  
Triggering is accomplished by an annular calorimeter with angular coverage above $20$ mrad
(e.g.\ 2 cm inner radius, 1 m downstream) by demanding three coincident hits carrying the beam energy.  
Signal events give rise to measurable impact parameters for the leading two tracks, and 
the excellent tracking provided by this design exploits this feature to reject background. 
Invariant mass reconstruction can provide an additional search
variable (see Sec.~\ref{subsec:expD}). 
More details are given in the text.   
{\bf Right:} \emph{Concentric purple contours:} Regions with detectable signal yield $\geq 10$ events, 
background rejection of $\sim 10^{-6}$ (yielding $S/B\gsim 1$), and an
impact parameter of at least 33 $\mu$m, 66$\mu$m, or $150 \mu$m, respectively, 
for the contours from the outside in. We assume a run time of $10^{6}$ s at $10^8$ e$^-$/s.  
\emph{Red Dotted Contour:} Analogous sensitivity with lower average current ($10^7$ e$^-$/s) and a smaller calorimeter aperture (10 mrad).
{\emph{Thin black dashed line:}}  a rough estimate of the total region of sensitivity that
could be accessible to this geometry using both displaced-vertex
discrimination and invariant mass search windows with good momentum resolution (see Sec.\ref{subsec:expD}).
\emph{Gray contours and Orange Stripe:} exclusions from past experiments (E137, E141, E774, 
  electron and muon anomalous magnetic moments, and $\Upsilon(3S)$ resonance searches) and the region that 
explains DAMA/LIBRA in a simple model --- see Figure \ref{fig:bigSummary} for more details. 
\label{fig:CollinearScenario}}
\end{figure*}

At even higher $\epsilon$ and lower masses, there exists the option of
halving the number of silicon strip tracking elements and placing them
directly into a defocused
primary electron beam of low intensity. For this study, we choose the
beam size to be about 1 cm $\times$ 1 cm and the beam energy to be 1
GeV. The beam intensity is limited by silicon occupancy to about
$10^{8}$ e$^-$/s, if we require occupancy of about 1\% in 1 cm
$\times$ 25 $\mu$m strips with a timing window of 20 -- 50 ns.

Triggering is again accomplished by a calorimeter, with a strategy
similar to case B and the same limitations.  For $A'$ masses of 20--50
MeV, decay opening angles $\sim 20-50$ mrad are anticipated, so the
calorimeter must extend close to the beam.  For simplicity we consider
an annular calorimeter with angular coverage above $20$ mrad (for
example, located at 2.5 meters from the target, with inner radius of 5
cm).  The beam electrons emerge from a 0.1 radiation-length tungsten
target in a Moli\`ere distribution, with typical transverse momenta of
5 MeV. Therefore less than $1\%$ of the electron beam hits the
calorimeter, leading to a $\lsim 1$ MHz singles rate, which is high
but manageable for a trigger requiring two hits.

With these parameters the $A'$ production rate is about 1 every ten
hours. Off-line track reconstruction can be used to remove the
backgrounds associated with the Coulomb scattering pile-up and other
background sources, in particular Bethe-Heitler pair production from
the target. The quality of the experiment will depend crucially on the
precision of the vertex reconstruction using the silicon strip
information.  Our sample point has typical impact parameter $\sim 160\,
\mu$m and laboratory decay lengths of order 2.3 mm, which should be
cleanly resolvable.  The sensitivity of this configuration, assuming
several different resolutions, is illustrated in Figure \ref{fig:CollinearScenario}.  

For smaller masses, the calorimeter must be placed at a narrower angle
or the beam energy reduced.  In either case, the Moli\`ere scattering
becomes more acute.  On the tails of the Moli\`ere distribution, one can
compensate by lowering the intensity of the beam.  At low beam
intensities, a fast scintillator/calorimeter trigger system will
resolve the passage of individual electrons in the beam 
(in a CW machine like CEBAF). Therefore, if the scintillator/calorimeter system
is segmented (e.g.~scintillating fiber calorimetry), the
trigger requirement can be simultaneous deposition of the beam energy
in more than one detection element --- typically three.  For larger
masses, the beam intensity would have to be increased, and the
silicon-strip occupancy presents a sharp barrier.

%%%%%%%%%%%%%%%%%%%%%%%%%%%%%%%%%
\subsection{High Resolution, High Rate Trident Spectrometer: $\epsilon =  3 \times 10^{-4}$; $m_{A'} = 1$ GeV}
\label{subsec:expD}

\begin{figure*}
\halfPage{\includegraphics[width=0.9\textwidth]{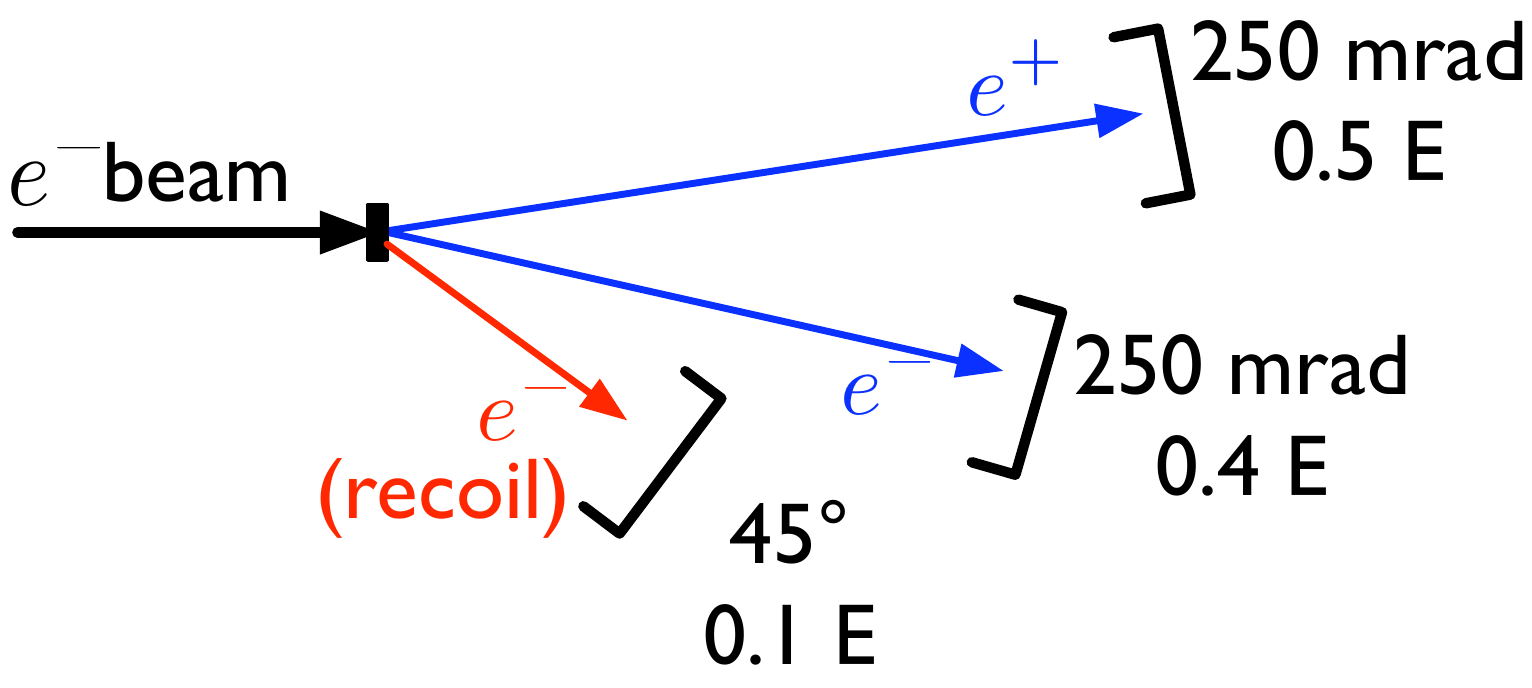}}
\halfPage{\includegraphics[width=0.9\textwidth]{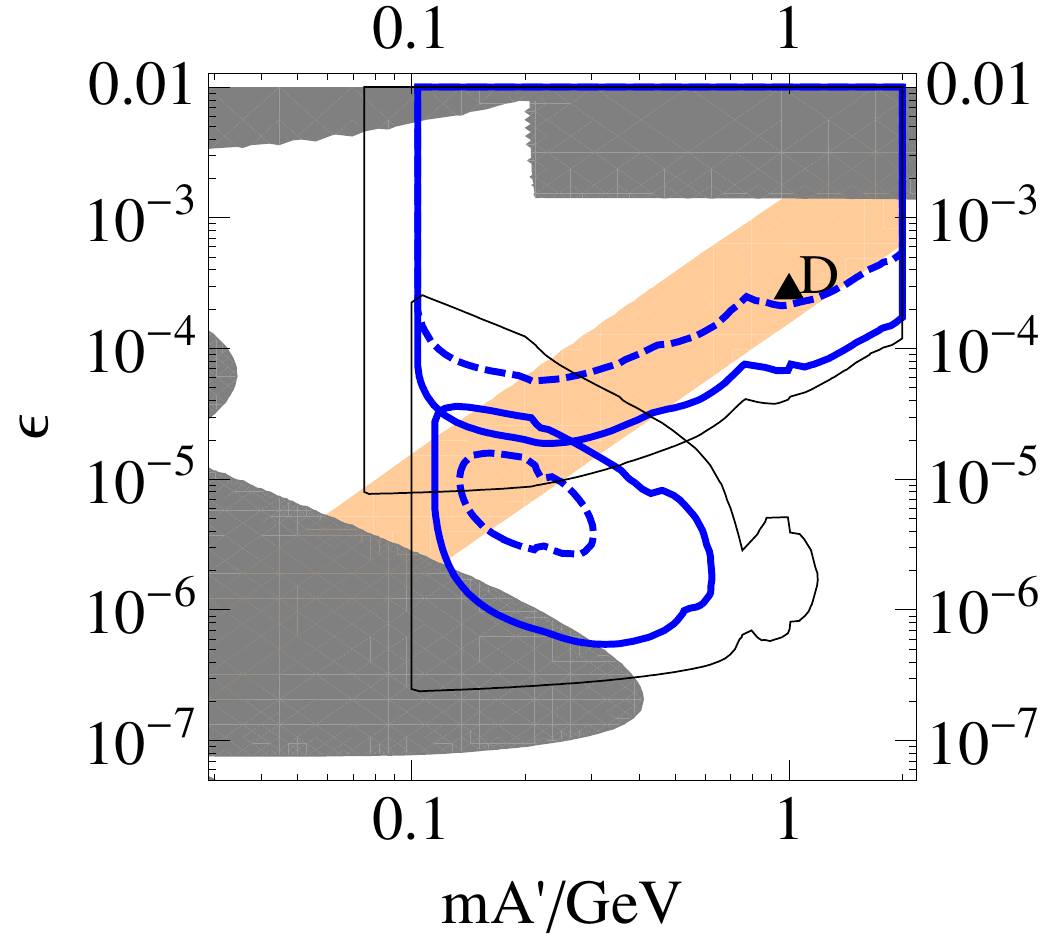}}
\caption{{\bf Left:} Schematic diagram of an experimental scenario for
  benchmark point D ($\epsilon\sim 3\times 10^{-4}$, $m_{A'}\sim 1$
  GeV).  An electron beam with an
  energy of $\sim 6$ GeV and a current of about 100 $\mu$A -- 200
  $\mu$A is incident upon a 0.1 radiation length aluminum target.
  A wide-angle high-resolution spectrometer allows triggering on
  events in which one electron and one positron carry most of the beam
  energy.  The signal is distinguished from background events with the
  help of various kinematic selection cuts (relatively symmetric $l^+l^-$
  final state and possible recoil electron tagging) and a ``bump hunt'' --- see
  text and appendix \ref{app:backgrounds} for further details.  
  {\bf Right:} Various estimates of the possible
  reaches of a wide-angle spectrometer, with (bottom) and without
  (top) tagging vertices displaced by $>1$ cm to reject background.
  In each case, the outer thin black line represents a significant
  total rate, with no geometric acceptance requirements
  ($S/\sqrt{B}>5$ in the no-vertex (top) region, 10 or more events in the
  vertex (bottom) region).  The thick blue curve shows the reach when
  decays are required to land more than 200 mrad away from the beam
  line, and the inner dotted curves assume an additional $1\%$ signal
  efficiency from acceptance.  In these two cases, each curve
  represents the total reach obtained by running at several beam
  energies.  
  %Obtaining good efficiency is evidently essential for the vertexing strategy to work.  
  \emph{Gray contours and Orange Stripe:} exclusions from past experiments (E137, E141, E774, 
  electron and muon anomalous magnetic moments, and $\Upsilon(3S)$ resonance searches) 
  and the region that explains DAMA/LIBRA in a simple model  --- 
see Figure \ref{fig:bigSummary} for more details. 
\label{fig:SpectrometerScenario}}
\end{figure*}

Large $A'$ masses present two challenges: a low production rate and
short $A'$ lifetime.  In the absence of a displaced vertex, the $A'$
can only be observed as a small peak on the electromagnetic trident
background. Reducing these backgrounds as much as possible is
essential here.  
Additionally, targets with somewhat lower $Z$ than tungsten are
preferable in this high $A'$ mass range in order to maintain 
charge coherence in scattering. 
For definiteness, we shall discuss the di-muon final
state, though it is arguable that the electron-positron final state is
preferable.  

As discussed in Section \ref{sec:production}, the trident background arises
from two subprocesses, which we call radiative and Bethe-Heitler
(c.f.~Figure \ref{fig:SigAndBkg}).  The radiative process gives an upper
bound on the ratio of signal to background as in equation (\ref{eq:IrreducibleBackground1}).  
The Bethe-Heitler process has a much larger ($\sim 100\times$) 
cross-section than the radiative trident process
due to collinear logarithmic enhancements in the 
$e\rightarrow e\,\gamma$ splitting and 
sub-process $\gamma\gamma \rightarrow \mu\mu$.
These enhancements can be avoided by demanding {\it kinematically 
symmetric} $\mu\mu$ decay products carrying the majority of the beam energy, 
and by demanding that the recoiling electron (if it can be identified)
scatter at a wide angle.
This preserves the large logarithm in the forward-peaked $A'$
production cross-section, while regulating all logs in the
Bethe-Heitler process.  These selections are discussed further in
Appendix \ref{app:backgrounds}.

In addition to the trident processes, radiation of real photons by
incident electrons, and their subsequent conversion in the target must
be considered. This process is naively enhanced by ${\cal
  O}(T/\alpha)$ relative to Bethe-Heitler trident production, but can
be rejected effectively with the same kinematic cuts.  It is, of
course, reducible by thinning the target, which allows a compensating
increase in average beam current.  We have not considered pile-up
processes, but assume they are small when the three products are
required to reproduce the beam energy within resolution.

For this scenario, we consider a 0.1-radiation-length aluminum
target in a 4 GeV beam.  The total yield of $A'$s is
roughly $10^{-16}$ per incident electron. If we assume an average beam
current of 250 $\mu$A (beam power of 1 MW) and an experimental
duration of $10^6$ sec, the total rate of $A'$ production is of order
one per second, or $\gsim 10^5$ per experiment.  These are emitted
in a cone of size $\sim 100$ mrad, with decay products at opening
angles near 250 mrad and the recoiling electron at a rather wide
angle, 0.5 radians.  The yield of
background tridents having a di-muon mass within one percent of the
$A'$ mass is, according to \eqref{eq:IrreducibleBackground1}, about
300 times larger, or $3 \times 10^7$ per experiment. The estimated cumulative sensitivity of this configuration, and similar ones 
obtained by lowering the beam energy down to $\sim 1 \GeV$,
 is illustrated in Figure \ref{fig:SpectrometerScenario}.
To obtain the contours in this figure, we require that $S/\sqrt{B}\ge 5$, i.e.~
$(S/\epsilon_b B_0) \times S \ge 25$, 
where $S$ is the signal rate, and $B$ is the background rate, $B_0$, times 
the background rejection efficiency $\epsilon_b$.  
We use equation (\ref{eq:IrreducibleBackground1}) to obtain $S/B_0$, and 
choose reasonable values for $\epsilon_b$.  

The signal rate above is, indeed, larger than necessary for
the $A'$ resonance to be statistically significant.  A less ambitious
(and perhaps more realistic) experiment would also suffice for
discovery.
There are at least three ways to back off from this scenario. One way
is evidently to improve the mass resolution. A second
way is to reduce the beam intensity, keeping the acceptance
complete. A reduction in beam current by a factor of 100 would still
leave a viable signal. The third way is to reduce the acceptance; a
one percent acceptance by itself would again leave a viable signal.

Optimization involves a choice of a combination of these
factors. 
Jefferson Laboratory looks like an especially appropriate
venue for this scenario, with two spectrometers with very good
electron momentum resolution.  In particular, the
small-acceptance, high-rate spectrometers in Hall A has momentum
resolution of order $10^{-4}$ and the large-acceptance Hall B
CLAS detector has electron momentum resolution better than $1\%$ \cite{Mecking:2003zu}. 
Therefore it would seem that using an electron-positron pair for the $A'$ decay products may
make more sense than using a di-muon pair.  However, we feel further
investigation is best done with the aid of expertise within the
Jefferson Laboratory experimental community.

\subsection*{High-Resolution Forward Spectrometers at Lower Luminosity }

\begin{figure}
\includegraphics[width=0.45\textwidth]{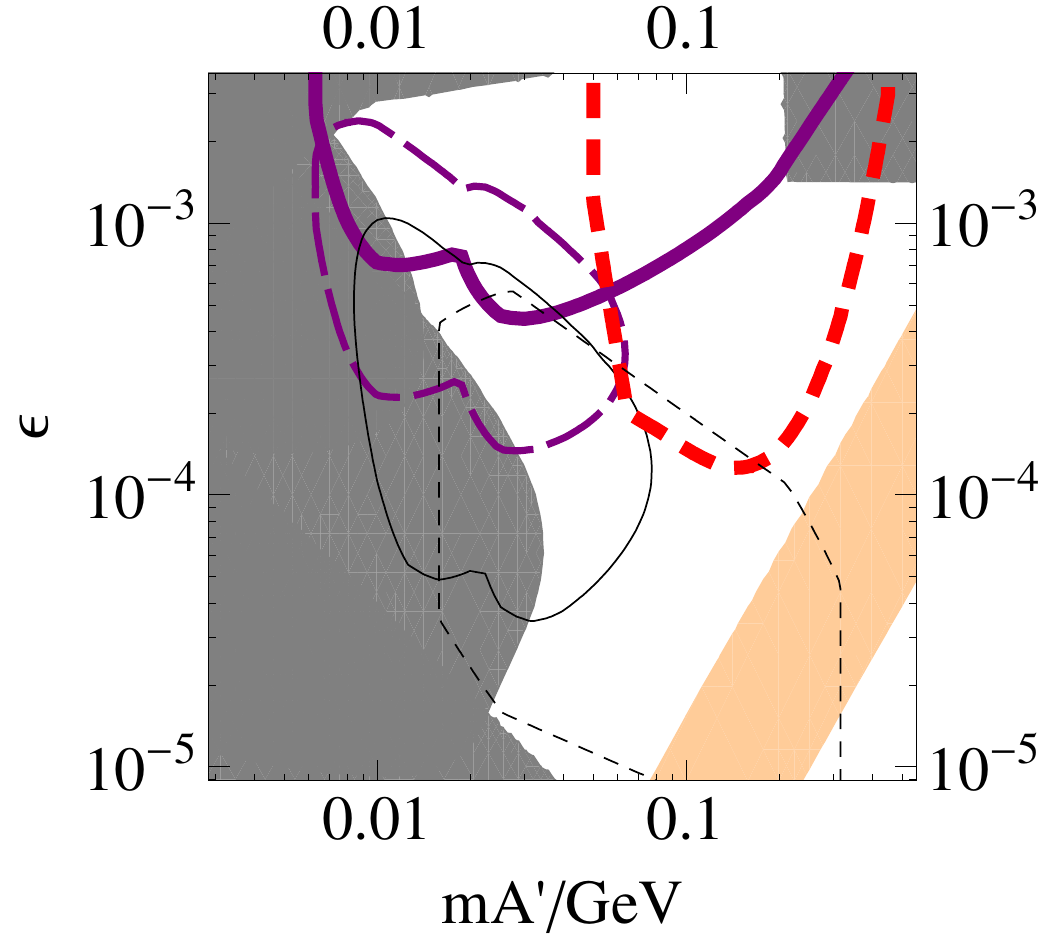}
\caption{\label{fig:forwardSpecNoV}{Estimated reach for the
  geometries of scenarios B (thick dashed red) and C (thick solid and
  long-dashed purple) in ``bump-hunt'' operation with upgraded mass resolution.
  For reference, the approximate expected reach of clean vertex-based
  searches discussed in sections B and C is given by the \emph{thin,
    black dashed contour} and the \emph{thin black solid contour}, respectively. \emph{Thick
    Dashed Red Contour:} The combined reach (with $S/\sqrt{B} \geq 5$) of the two
  two-arm spectrometer geometries introduced in Section \ref{subsec:expB} and Figure
  \ref{fig:ForwSpec} with a 6 GeV beam. 
  \emph{Thick Solid Purple
    Contour:} $S/\sqrt{B}\geq 5$ for the combined reach of the different ``diffuse beam'' 
    scenarios in Section \ref{subsec:expC}, assuming $1\%$ mass resolution.
  \emph{Long Dashed, Thinner Purple Contour:} Same as the thick solid purple contour, 
  but using vertexing to reduce backgrounds, with an assumed ``optimistic'' rejection of $10^{-2}$ 
  for an impact parameter cut of 10 $\mu$m applied to the lepton pair. 
  \emph{Gray contours and Orange Stripe:} exclusions from past experiments (E137, E141, E774, 
  electron and muon anomalous magnetic moments, and $\Upsilon(3S)$ resonance searches)  and the 
 region that explains DAMA/LIBRA in a simple model  --- 
see Figure \ref{fig:bigSummary} for more details.
}}
\end{figure}

As we lower $m_{A'}$ while keeping $\epsilon$ large, the production angles
may become too low for big spectrometers such as those found in Hall A at JLab.  
The geometries  of the detectors sketched in Sections \ref{subsec:expB} and 
\ref{subsec:expC} may be more appropriate.
The growth of the production cross-section as $1/m_{A'}^2$ compensates for
the lower average currents demanded by these scenarios (10-100 nA for the
two-arm spectrometer B (described in Section \ref{subsec:expB}) and $10^{8}$ e$^-$/s  
when tracking planes are
inserted in the beam as in C (described in Section \ref{subsec:expC})).  

In scenarios B and C, in the interest of simplicity, we did not assume a high 
mass resolution for the spectrometers. However, a high mass resolution, 
e.g.~via addition of magnetic fields, is not at all impractical in principle. 
Assuming a 1\% mass resolution, the potential sensitivity of these modified 
scenarios is sketched in Figure \ref{fig:forwardSpecNoV}.  
Again, it will be important to detect the 
wider-angle recoil electron in order to reduce backgrounds. 
This requires increasing the tracking and calorimeter coverage to around 250 mrad 
for the two-arm spectrometer B as well as for the diffuse-beam, collinear
 configuration C.

The gap between the spectrometer and resolved vertex regimes coincides
with $A'$ decay lengths of order the $1\sigma$ vertex resolution of a given detector. 
In this regime, one can gain sensitivity by imposing a loose impact
parameter/vertex requirement that reduces background rates by a few 
orders of magnitude.  Clearly, this strategy is plagued by the
difficulties of both the spectrometry and vertexing approaches, and
would only be attempted in a later stage when the detector and
backgrounds are well understood.  We do not discuss it further.
  
%%%%%%%%%%%%%%%%%%%%%%%%%%%%%%%%%%%%%%%
\subsection{High power, Low Energy Beam Dump; $\epsilon = 5\times 10^{-8}$; $m_{A'} = 50$ MeV}
\label{subsec:expE}

\begin{figure*}
\begin{minipage}{0.49\textwidth}
\begin{center}
\includegraphics[width=0.7\textwidth]{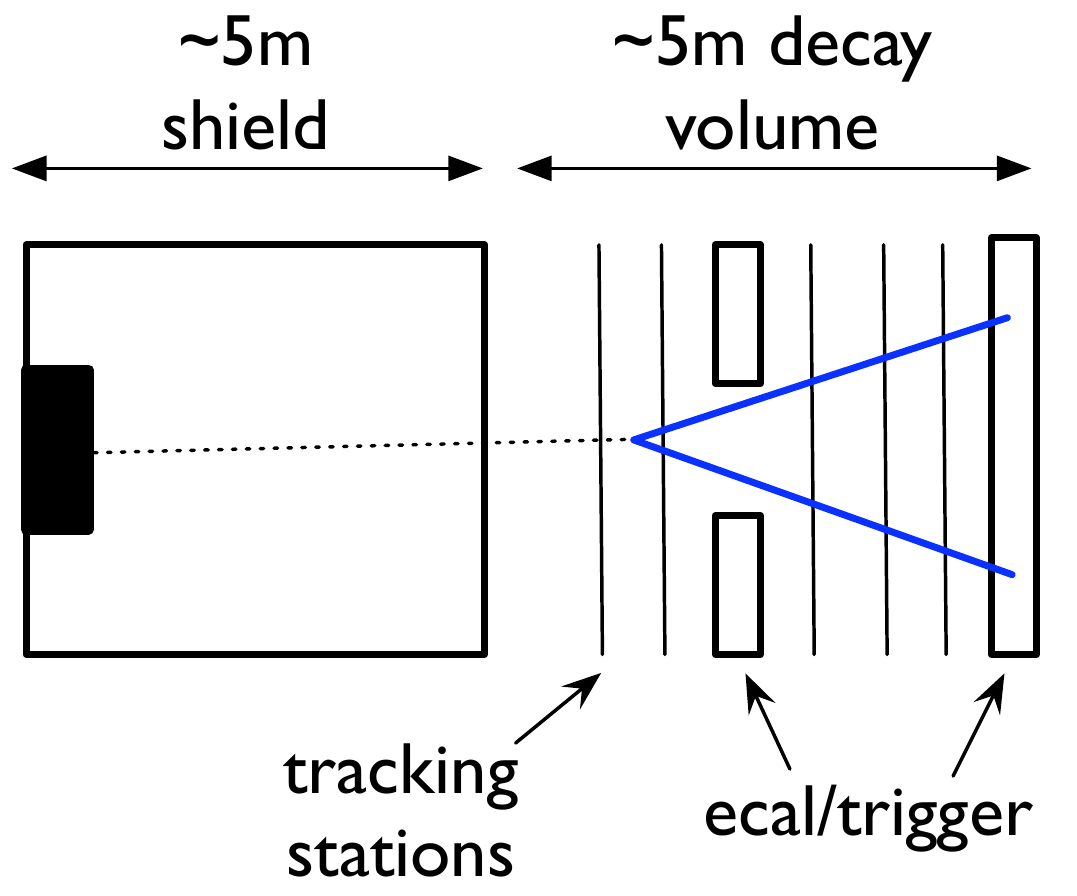}\;\;
\end{center}
\end{minipage}
\begin{minipage}{0.49\textwidth}
\includegraphics[width=1.0\textwidth]{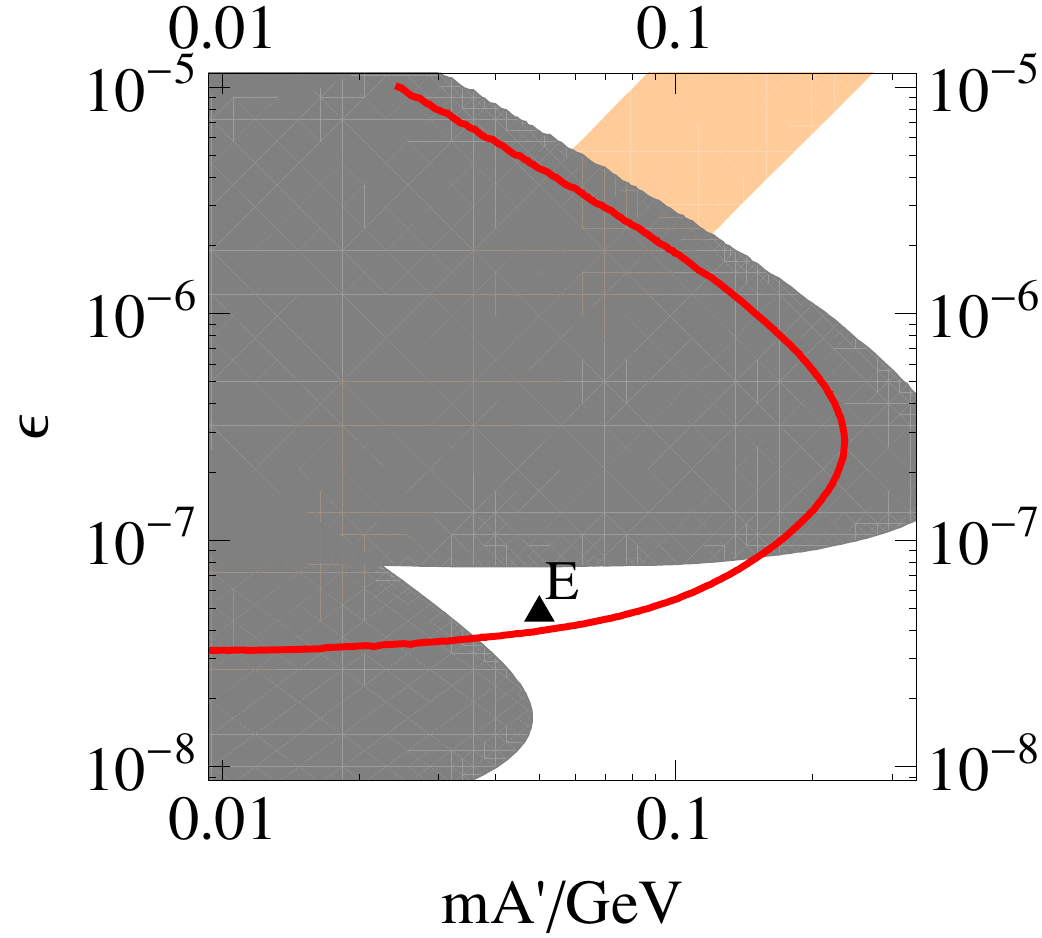}
\end{minipage}
\caption{{\bf Left:}  Schematic diagram of a beam dump design for benchmark point E 
($\epsilon\sim 5\times 10^{-8}$, $m_{A'}\sim 50$ MeV).  
A 200 MeV  electron beam with a large current of about 5 mA (delivering 1 megawatt in power) 
is incident upon a thick tungsten target that 
together with shielding is about 5 m in length.  Behind the shielding is a decay region 5 m 
long, consisting of a tracking system (2 m $\times$ 2 m transverse to the beam line) and surrounded 
by electromagnetic calorimeters --- see text for further details.  
{\bf Right:} 
\emph{Solid Red Contour:} 10 events with $A'$ energies above 100 MeV 
after the experiment has run for $10^6$ s (5000 C total charge dumped).  
 \emph{Gray contours and Orange Stripe:} exclusions from past experiments (E137 and SN1987A)
 and the region that explains DAMA/LIBRA in a simple model  --- 
see Figure \ref{fig:bigSummary} for more details. 
\label{fig:HPDumpScenario}}
\end{figure*}

Values of $\epsilon$ below the  E137 limit require very intense
beams simply to produce enough $A'$s to detect. Beam power limitations
force one downward in beam energy. We choose a 200 MeV beam of
electrons with an average current of 5 mA, representing a beam power
of 1 MW. The  $A'$ production rate is about $2\times10^{-19}$ per electron
dumped. The laboratory decay length is about 2.5 km. The divergence of
the $A'$ beam is about 100 mrad. We consider a decay region 5 m
long, with its front end located 5 meters downstream of the dump. A
tracking system, perhaps in the style of E137, with transverse
dimensions 2 m $\times$ 2 m is distributed throughout the decay volume to
capture the decay products from the $A'$ decays. It is surrounded by
electromagnetic calorimetry designed to efficiently capture the
electrons and positrons emergent from the tracking volume (see Figure 
\ref{fig:HPDumpScenario}). With these parameters, the yield of detected $A'$s is
marginal --- about 5--10 per $10^6$ seconds.

This scenario has not been optimized, and other versions less awkward
can be contemplated. However, the region of the exclusion plot that is
covered by any such experiment will be modest. Therefore a real design
is likely to be opportunistic. If the experiment can be run
parasitically for a long period of time, the benefit-to-cost ratio may
rise sufficiently high to make such an effort attractive.

If one contemplates utilizing a higher energy dump of 1 MW power, then
the number of electrons dumped decreases and, keeping $m_{A'}$ and
$\epsilon$ fixed, the yield of $A'$s decreases as well. The region of
sensitivity at $A'$ masses of about 200 MeV should, to a fair
approximation, merge with the reach of E137, as estimated in
Figure \ref{fig:HPDumpScenario}.

%%%%%%%%%%%%%%%%%%
%%%%%%%%%%%%%%%%%%
%\newpage

\section{Discussion and Conclusions}\label{sec:discussion}

In this paper, we have described five scenarios for fixed-target
experiments that probe kinetically mixed $U(1)$'s with MeV-GeV masses.
Kinetic mixing of size $\epsilon \sim 10^{-2} -10^{-8}$ between a light gauge boson 
and the photon can be generated by loops of particles at any mass scale, with the
magnitude determined by the structure of high-scale physics.  
An MeV-GeV mass for the $A'$ can in turn be generated from the weak scale, 
especially in a supersymmetric context.  An $A'$ in this mass range is also of
interest as a possible explanation of several current dark matter
anomalies.

The parameter space of $A'$ mass $m_{A'}$ and mixing $\epsilon$ has been
constrained from two corners by existing data.  Beam dump experiments
and supernovas exclude the low-mass, small-$\epsilon$ region.
Larger $\epsilon \sim 10^{-3}-10^{-2}$ are constrained for a broad
range of masses by lepton anomalous magnetic moments and B-factory
searches.  

The five approaches we have described cover the remaining parameter
space using fixed-target experiments of various geometries and 200 MeV--6 GeV beams.  
A natural extension of past beam dumps, with modest intensity and 10
cm--1 m length, can fill in the crevice of parameter space between past
beam dumps.  Beam dumps are not well suited to searching for $A'$ with
less displaced decays.  For these parameter ranges, thin-target
experiments are required.

Any thin-target experiment must contend with the backgrounds
from electromagnetic electron scattering and trident production, which
can be tackled with a combination of kinematics and displaced vertex
selection.  Depending on $m_{A'}$, more forward or wide-angle
geometries are called for, and small-scale silicon micro-strip tracking 
can be utilized to isolate displaced decays.  
We have considered three such scenarios: a
forward two-arm spectrometer, a collinear detector in a diffuse,
low-intensity beam, and a wide-angle spectrometer.  Together, they are
sensitive in the range $\epsilon \sim 10^{-5}-10^{-3}$ for $A'$ masses
from 10 MeV to 1 GeV.

The wide-angle scenario is of particular interest, because existing
spectrometers can cover a large fraction of its
reach.  The Hall A spectrometers and the CLAS detector \cite{Mecking:2003zu} at JLab 
seem well suited for initial searches, and other labs may have 
comparable capabilities.

Searches at low $\epsilon$, below the reach of the dump experiment
E137, are limited by practical rate limitations.  Power above a
megawatt (MW) is difficult to sustain, making $\epsilon \sim
10^{-8}-10^{-7}$ inaccessible with beams of any energy in under a year
of running.  Our fifth scenario saturates this limit, with a 200 MeV
MW dump, which can possibly be accommodated at the JLab Free-Electron Laser accelerator.

When combined with existing limits, these five scenarios can either
confirm the existence of new $U(1)$ gauge forces at low masses or
close the door on their most likely parameter range.

For masses below the electron threshold, very different experimental techniques 
are called for. These have been developed in \cite{Afanasev:2008fv,Afanasev:2008jt}.

We have restricted our discussion to the simplest scenario: a single $U(1)$ gauge boson
that decays directly to electrons. 
In a larger ``dark sector'', somewhere between
this minimal scenario and the full complexity of Standard Model
physics, decays within the dark sector dominate \cite{Essig:2009nc}.  
These dark-sector cascades can return some or all of the $A'$ 
energy to Standard Model-charged particles, with lifetimes 
controlled by $\epsilon^{-2}$ for vector bosons 
and much longer lifetimes $\propto \epsilon^{-4}$ for scalars.  The 
limits and reaches discussed here apply directly to any spin-1
bosons in the dark sector that decay directly to a lepton pair.  
These experiments are
also sensitive to dark sector cascades involving spin-0 states, 
with appropriately deformed exclusion regions not discussed here.
Besides frameworks with kinetically mixed $U(1)$, these experiments are sensitive to direct production of light (pseudo) scalars 
(e.g. \cite{Hooper:2009gm,Mardon:2009gw,Nomura:2008ru,Strassler:2006im}).
It is likely that related designs more optimally cover these scenarios.

We have focused here on experimental approaches tailored to $A'$ searches
in electron beams, but analyses in this spirit may be possible with existing data, 
for example by using beam-halo impacts in collider experiments or neutrino 
production beams and detectors such as those at Fermilab and KEK.  
We also have not explored the potential of muon-beam experiments, 
which may be ideal for searches for $A'$s above the muon mass, which are produced with
rates comparable to those for an electron beam, but with much lower
electromagnetic backgrounds.

%%%%%%%%%%%%%%%%%%
%%%%%%%%%%%%%%%%%%
\acknowledgments

We thank Andrei Afanasev, Dan Dale, Alex Dzierba, and Richard Jones
for information regarding opportunities at JLab for implementing this
program.  We are grateful to John Cumalat, Mathew Graham, John Jaros,
Aaron Roodman and Jay Wacker for feedback and discussion regarding our
experimental scenarios, and to Savas Dimopoulos, Michael Peskin, and
Matt Reece for many fruitful discussions. We especially thank Takashi
Maruyama for providing us with background estimates for the forward
two-arm spectrometer design.  RE and PS are supported by the US DOE
under contract number DE-AC02-76SF00515.

%%%%%%%%%%%%%%%%%%
%%%%%%%%%%%%%%%%%%
\appendix 

%%%%%%%%%%%%%%%%%%
\section{$A'$ Production Formulas}\label{app:Details}
In this appendix, we first present the cross-section for the production of the massive $U(1)'$ 
``dark photon'', $A'$, by initial- or final-state
radiation off a \emph{single} electron hitting a fixed target of atomic number $Z$.  
This process is
analogous to photon bremsstrahlung, except that the coupling of the $A'$ to 
electrons is $\epsilon \cdot e$ and the $A'$ mass, $m_{A'}$, is much larger than 
the electron mass, $m_e$, which
significantly alters both the kinematics and the rate of the process.
The qualitative behavior has already been summarized in Section
\ref{sec:production}.  

We want to calculate the $A'$-production cross-section
\be\label{eq:CrossSection}
\f{d\sigma(e(p)+Z(P_i) \rightarrow e(p') + A'(k) + Z(P_f))}{d E_{A'}
  d\cos\theta_{A'}}
\ee
in the Weizs\"acker-Williams approximation following \cite{Kim:1973he,Tsai:1973py,Tsai:1986tx}, 
where $k=(E_{A'}, \vec{k})$ is the momentum of the outgoing $A'$, 
$\theta_{A'}$ is the angle of
its momentum relative to the incoming electron momentum $\vec p$ in
the lab frame, $p=(E_0,\vec p)$ and $P_i = (M_i,\vec 0)$ are the
initial momenta of the electron and the target of mass $M_i$ and
atomic number $Z$, and $p'=(E',\vec{p}')$ and $P_f$ are the outgoing four-momenta of the
electron and target, which are integrated over.  

In the frame of the incoming electron, the rapidly moving atom sources
a cloud of effective photons, off which the electron scatters to
radiate an $A'$.  Though these photons are spacelike, their virtuality
is small compared to other invariants in the problem (for example
$m_{A'}$), so that the interaction of the electron with the target is
dominated by transverse polarizations. Therefore, it is related to the cross-section for real-photon
scattering, $e(p) \gamma(q) \rightarrow e(p') A'(k)$ with $q=P_i -
P_f$ by  \cite{Tsai:1986tx} 
\bea
&&\f{d\sigma(p+P_i \rightarrow p' + k + P_f)}{d E_{A'}
  d\cos \theta_{A'}}  = \left(\f{\alpha \chi}{\pi}\right) \left(\f{E_0 x
  \beta_{A'}}{(1-x)}\right) \qquad \nonumber \\
  && \qquad\qquad\qquad \times \f{d\sigma(p + q \rightarrow p' + k)}{d(p\cdot k)}\bigg|_{t=t_{min}},
\eea 
where 
\bea
x&\equiv& E_{A'}/E_0, \nonumber \\
t&\equiv& -q^2.
\eea
We specify the kinematics at $t=t_{min}$ and the effective photon flux
$\f{\alpha\chi}{\pi}$ below.  Note that $t$ is \emph{not} one of the 
Mandelstam variables for the $2\to2$ process, which will be denoted by $t_2$ --- see below.   

For a given $A'$ four-momentum $k$, the
virtuality $t$ has its minimum value $t_{min}$ when 
$\vec q$ is collinear with the three-vector $\vec k -\vec{p}$.  
Solving the mass-shell conditions $p'^2 = (q + p - k)^2 =
m_e^2$ and $P_f^2 = (P_i - q)^2 = M_i^2$ with the collinear geometry,
and keeping only leading effects in
\be
\f{m_{A'}^2}{E_k^2},\quad \f{m_e^2}{E'^2}, \quad \theta_{A'},
\quad \f{|\vec{q}|}{E'},
\ee
(with $|\vec{q}|$ defined below), we find 
\bea
&& q^0= |\vec{q}|^2/2 M_i \approx 0 ,\quad |\vec{q}| = \frac{U}{2 E_0 (1-x)}, \\
&& t_{min} = -q_{min}^2 \approx \left(\frac{U}{2 E_0 (1-x)}\right)^2,
\eea
where 
\bea
%x & = & E_{A'}/E_1, \\
U & \equiv & U(x,\theta_{A'}) = E_0^2 \theta_{A'}^2 x + m_{A'}^2 \f{1-x}{x} + m_e^2 x.
\eea
At this kinematics,
\bea
- \tilde u &\equiv& m_e^2 -u_2  = 2 p \cdot k - m_{A'}^2 = U, \\
  \tilde s &\equiv& s_2 - m_e^2 = 2 p' \cdot k  + m_{A'}^2 = \f{U}{1-x}, \\
         t _2& =&  (p-p')^2 = - \f{U x}{1-x} + m_{A'}^2,
\eea
where $s_2$, $t_2$, and $u_2$ are the Mandelstam variables for the $2\to 2$ process.
The cross-section for the 
$2\rightarrow 2$ process is therefore
\begin{widetext}
\bea
\f{d\sigma}{d(p\cdot k)} = 2 \f{d\sigma}{d t_2} 
%= 2 \f{1}{64\pi s {\vec q}^2} |M|^2 
\approx \f{1}{8\pi (s_2-m_e^2)^2} |{\cal M}|^2
 &=& \f{4\pi\alpha^2\epsilon^2}{\tilde s^2}\left(
\f{\tilde s}{- \tilde u} + \f{- \tilde u}{\tilde
  s}+\f{2 m_{A'}^2 t_2}{-\tilde u \tilde s} \right) \nonumber \\
 &=&  (4\pi\alpha^2\epsilon^2) \f{(1-x)}{U^2} \left[
1+(1-x)^2+\f{2 (1-x)^2 m_{A'}^2}{U^2} \Big(m_{A'}^2 - \f{U x}{1-x}\Big) \right],
\label{e3}
\eea
\end{widetext}
where we have dropped the $t$-dependence of $\f{d\sigma}{dt_2}$ and
terms of order $m_e^2$ in $|{\cal M}|^2$.  
Therefore, the Weizs\"acker-Williams approximation to the 
cross-section \ref{eq:CrossSection} is given by
\begin{widetext}
\be
%\f{1}{E_1 x} \f{d\sigma_{3\rightarrow 2}}{d E_{A'}
%  d\cos \theta_{A'}} =  
\f{1}{E_0^2 x} \f{d\sigma_{3\rightarrow 2}}{d x \,d \cos\theta_{A'}}
= (8 \alpha^3\epsilon^2 \chi \beta_{A'}) \left[
\f{1-x+\f{x^2}{2}}{U^2}+ \f{(1-x)^2 m_{A'}^2}{U^4} \Big(m_{A'}^2 - \f{U x}{1-x}\Big) \right],\label{almostWW}
\ee
\end{widetext}
where $\beta_{A'}\equiv \sqrt{1-m_{A'}^2/E_0^2}$.   
The $x$-differential cross-section is obtained by integrating
\eqref{almostWW} with respect to $\theta_{A'}$ 
%from $U=U(x,0)$ to $\infty$.  
(we will see below that $\chi$ actually depends on $\theta_{A'}$, but 
this can be neglected to excellent approximation).  
The first term in square brackets integrates to
\be
\f{1-x+\f{x^2}{2}}{U(x,\theta_{A'}=0)}.
\ee
In the limit $m_{A'} \rightarrow 0$, this becomes the standard photon 
bremsstrahlung result with a $\f{1}{x}$ singularity, while the second
term in the square brackets vanishes.  However, finite $m_{A'}$ regulates
this singularity, and in the case of
interest, namely $m_{A'} \gg m_e$, we have $U(x,0) \approx m_{A'}^2 \f{1-x}{x}$. 
The second term integrates to $-\f{x^2}{6 U(x,0)} + {\cal
  O}({m_e^2}{U^2})$,  so that 
\bea
\f{d\sigma_{3\rightarrow 2}}{d x} = && (8 \alpha^3\epsilon^2 \chi
\beta_{A'})
\left( m_{A'}^2 \f{1-x}{x} + m_e^2 x \right)^{-1} \nonumber \\
&& \times \Big(1-x+\f{x^2}{3}\Big).
\eea
This has an approximate \emph{soft electron} singularity, regulated by
the electron mass at $(1-x)_{c1} = \f{m_e^2}{m_{A'}^2}$.  Though not
explicit in this formula, our approximations also break down if the
electron energy $(1-x) E_0 \lesssim |\vec q|$; this also regulates the
cross-section, cutting off $\log(1-x)$ at  $(1-x)_{c2} =
\f{m_{A'}^2}{E_0^2}$.  Since one cutoff or the other is always larger
than their geometric mean $m_e/E_0$, the $A'$ is always produced 
from a relativistic electron.  The $x$-integrated cross-section is therefore 
\bea
&&\sigma \approx \f{8}{3} \f{\alpha^3\epsilon^2 \beta_{A'}}{m_{A'}^2}\, \chi 
\log\left(\f{1}{(1-x)_c}\right), \\
&&(1-x)_c = \max\bigg( \f{m_e^2}{m_{A'}^2},\, \f{m_{A'}^2}{E_0^2}\bigg).
\label{sigmaApp}
\eea

As we have noted in Section \ref{sec:production}, the characteristic angle of 
$A'$ emission is set by $U(x,\theta_{A'}) - U(x,0) \sim U(x,0)$, so 
$\theta_{A'} \sim \f{m_{A'}\sqrt{\overline{1-x}}}{E_0}$, where the median 
value of $1-x$ is $\overline{1-x} \sim \max\left(\f{m_e}{m_{A'}},\f{m_{A'}}{E_0}\right)$.  
This is parametrically smaller than the angle of the $A'$ decay products with 
respect to the incoming electron, namely $\sim m_{A'}/E_0$.

\vskip 3mm

We turn next to the definition of $\chi$, which is an effective flux
of photons integrated from $t=t_{min}$ to $t_{max}$, the total
center-of-mass energy of the collision.  We refer the reader to 
\cite{Kim:1973he,Tsai:1973py} for more details.  

For a general electric form factor $G_2(t)$,
\be
\chi \equiv \int_{t_{min}}^{t_{max}}  dt \f{t-t_{min}}{t^2} G_2(t).\label{ChiExp}
\ee
(We note that the other form factor, $G_1(t)$, contributes only a negligible amount in 
all cases of interest.)
Although the virtual photon propagator
squared, $1/t^2$, is dominated at $t=t_{min}$, the final-state phase space is 
proportional to $dt (t-t_{min})$, so that virtual photons at all
scales contribute to $A'$ production.  
As discussed in \cite{Kim:1973he,Tsai:1973py},
the physical upper bound may be set not by the center-of-mass energy, but by $t_{max} \sim
m_{A'}^2$, at which the full $2\rightarrow 3$ matrix element begins to shut
off.  

For most energies in question, $G_2(t)$ is dominated by an elastic component 
\be
G_{2,el}(t)= \left(\f{a^2 t}{1+a^2 t} \right)^2
\left(\f{1}{1+t/d} \right)^2 Z^2,
\ee
where the first term parametrizes electron screening (the elastic atomic form factor) 
with $a=111\,Z^{-1/3}/m_e$, and the second finite nuclear size (the elastic nuclear form 
factor) with $d=0.164 \mbox{ GeV}^2 A^{-2/3}$.  
We have multiplied together the simple parametrizations used for each in \cite{Kim:1973he}.  
The logarithm from integrating \eqref{ChiExp} is large for 
$t_{min} < d$, which is true for most of the range of interest.  
However, for heavy $A'$, the elastic contribution is suppressed and is 
comparable to an inelastic term,
\be
G_{2, in}(t)= \left(\f{a'^2 t}{1+a'^2 t} \right)^2 \left(\f{1+\f{t}{4
    m_p^2} (\mu_p^2-1)}{(1+\f{t}{0.71\,{\rm GeV}^2})^4} \right)^2 Z,
\ee
where the first term parametrizes the inelastic atomic form factor and the second 
the inelastic nuclear form factor, and where $a'=773 \,Z^{-2/3}/m_e$, $m_p$ is the proton 
mass, and $\mu_p=2.79$ \cite{Kim:1973he}. 
This expression is valid when $t/4m_p^2$ is small, which is the case for $m_{A'}$ 
in the range of interest in this paper.  One can show that the contribution 
from the other inelastic nuclear form factor $G_1(t)$ is negligible.   

\begin{figure}
\includegraphics[width=0.45\textwidth]{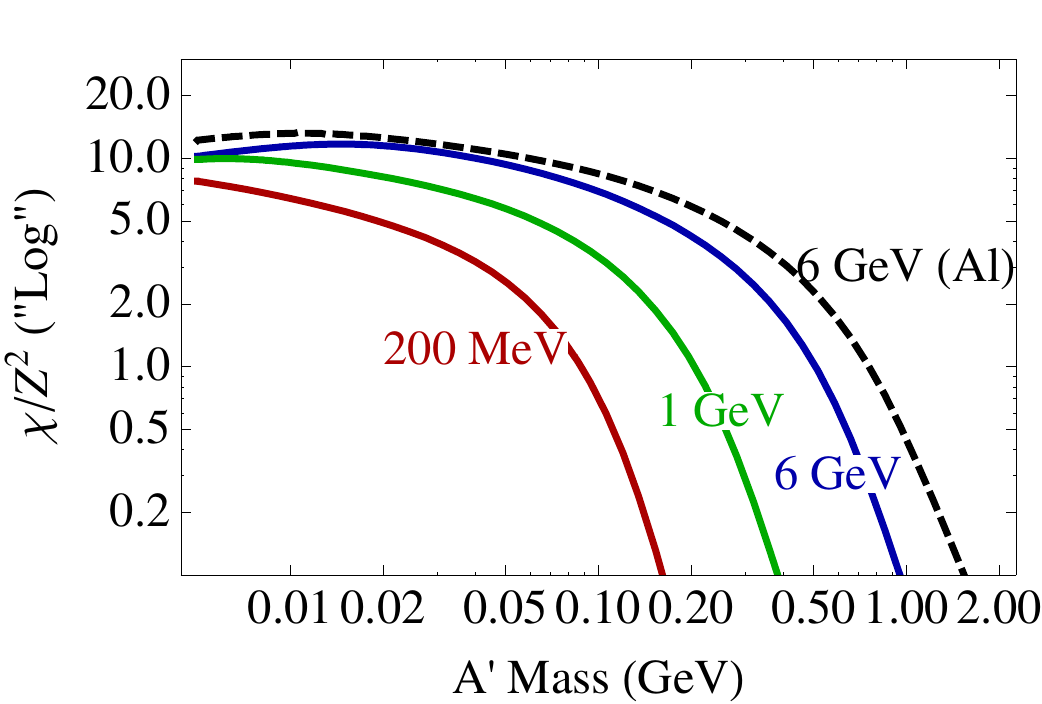}
\caption{\label{fig:chiLOG} The factor ${\cal L} og = \chi/Z^2$
  appearing in equations \eqref{eq:dSigmadxdcos} and \eqref{dSigmadX} in 
  Section \ref{sec:production}. Solid
  lines from left to right correspond to beams of energy 200 MeV, 1 GeV,
  and 6 GeV, respectively, incident on a Tungsten target.  The dashed line
  corresponds to a 6 GeV beam incident on an Aluminum target.}
\end{figure}

At high masses, these simple
parameterizations of the form factors are uncertain at the
order-of-magnitude level.  Using $G_{2,el}+G_{2,in}$ in \eqref{ChiExp},
and setting $t_{min}=(m_{A'}^2/2E_0)^2$, $t_{max}=m_{A'}^2$, we obtain
$\chi/Z^2$ shown in Figure \ref{fig:chiLOG}.  

%%%%%%%%%%%%%%%%%%%%%%%%%%%%%%%%%%
\section{Production From Thin and Thick Targets of $A'$s with Finite Lifetime}\label{app:moreproduction}
Consider an electron beam with energy $E_0$ incident 
on a target.
The total number of $A'$ that get produced in the target with energy 
$E_{A'} \equiv x E_0$ and decay at a distance $z$ behind 
the front edge of the target is given by
\bea
\f{dN}{dx dz} = &&
N_e \f{N_0 X_0}{A}\,
\int_{E_{A'}}^{E_0}  dE_1 \int_0^T  dt\, I(E_1; E_0, t) \nonumber \\
&&\times \left(\f{E_0}{E_1} \f{d\sigma}{d x'}\right)_{x'=\f{E_{A'}}{E_1}}
\, \f{dP(z-\f{X_0}{\rho}\,t)}{dz},\label{dNdxdz}
\eea
where $E_0$ is the incident energy, $N_e$ the number of incident
electrons, $N_0=6.02 \times 10^{23}$ mole$^{-1}$, $\rho$ and $X_0$ are the 
density (in g/cm$^3$) and unit radiation length (in g/cm$^2$) of the
target material, respectively, and 
\be
\f{dP(\ell)}{d\ell} = \f{1}{\ell_0} e^{-\ell/\ell_0}
\ee
is the differential decay probability where $\ell_0 \equiv \gamma c \tau$ 
$=E_1 c\tau /m_{A'}$ 
is the $A'$ decay length given in \eqref{gammaCTau}.  
Also, 
\be
I(E_1; E_0, t) \approx 
\begin{cases}
\f{1}{E_0} y^{b t- 1} \,b t &  T \gtrsim 1 \\
\delta(E_{1}-E_0) &  T \ll 1 
\end{cases}
\ee
is the energy distribution of electrons at position $t$ in the
target, where $y \equiv \f{E_0-E_1}{E_0}$ and $b=4/3$.

We can perform the $t$-integration explicitly in the limit of a very
thin or thick target ($T\ll 1$ or $T \gg 1$). For a thin target, we find 
\be
\f{dN_{thin}}{dx dz} = N_e \f{N_0 \rho \ell_0}{A}\, \f{d\sigma}{dx} 
\left( e^{\f{T X_0}{\rho \ell_0}}-1 \right) \f{dP(z)}{dz}.
\ee
For a thick target, we neglect the $t$-dependence in the $A'$ decay probability in
\eqref{dNdxdz}, since most production occurs within the first radiation
length and thus well before the end of the dump.  Here we find
\bea
\f{dN_{thick}}{dx dz}  \approx  && N_e \f{N_0 X_0}{A}\,
\int_{E_{A'}}^{E_0}  dE_1\, \tilde I(E_1; E_0, T) \nonumber \\
&& \ \ \times \left(\f{E_0}{E_1} \f{d\sigma}{d x'}\right)_{x'=\f{E_{A'}}{E_1}}\,
\f{d P(z)}{dz},
\eea
where
\bea
&& \tilde I(E_1; E_0, T)  =  \int_0^T  dt\, I(E_1; E_0, t) \nonumber \\
 && \qquad \approx 
\f{1 + y^{b T} (b T \ln y  - 1)}{E_0 b y (\ln y)^2 }
\rightarrow \f{1}{E_0 b y  (\ln y)^2}
\eea
as $T \rightarrow \infty$. For finite $T$, the limiting form is a good
approximation for small and moderate $y$, i.e. for electrons that carry a
large fraction of the initial beam energy (for $y<0.5$ and $T>7$ it is correct
to within $1\%$).

We note that by \eqref{dSigmadX} and \eqref{dNdxdz}, $\f{1}{N_e}
\f{dN}{dxdz}$ is proportional to 
\be \f{1}{\ell_0} \f{8 \alpha^3\epsilon^2 Z^2
  \chi}{m_{A'}^2} \f{N_0 X_0}{A}.
\ee
However, $\f{1}{X_0} = \f{4\alpha^3 N_0}{m_e^2 A} [Z^2 (L_{rad} -f(Z)) +Z
  L'_{rad}]$, where $L_{rad}$, $L'_{rad}$, and $f$ are logs set by the
atomic form factors of the target atoms \cite{Amsler:2008zzb}, and we obtain   
 \be
\f{dN}{dxdz} \sim \f{\min(T,1)}{\ell_0} \f{m_e^2}{m_{A'}^2} \epsilon^2,
\ee 
with only logarithmic dependence on the target nucleus $Z$.   
This expression has a simple physical interpretation: an electron is
slowed in a radiation length by radiating a small number of relatively
hard but collinear photons.  The probability of instead radiating an
$A'$ is suppressed by the squared ratio of the couplings, $\epsilon^2$, 
and the squared ratio of the masses $\f{m_e^2}{m_{A'}^2}$, because it 
requires a higher invariant-mass intermediate state.  

For decays at a given $z$, we applied an acceptance based on the detector
geometry and energy cuts before integrating to obtain the expected total rate.

%%%%%%%%%%%%%%%%%%%%%%%%%%%%%%%%%%
\section{Kinematics of Signal and Controlling Backgrounds}\label{app:backgrounds}

The dominant QED backgrounds for $A'$ production are the trident
reactions shown in Figure \ref{fig:SigAndBkg}. The $\gamma^*$ process
contributes an irreducible background to $A'\rightarrow l^+l^-$.  The
Bethe-Heitler process has a much larger rate, but can be controlled by
exploiting its very different kinematics compared to the signal.  Our
aim in this appendix is to quantitatively describe the singularity
structure of the Bethe-Heitler process and derive an effective set of cuts on
lab-frame observables.  Of course a more accurate basis for a final
design would rely on monte carlo for these processes, but the simple
calculation clarifies the origin of the large Bethe-Heitler
cross-section and how to regulate it by cutting away from the
dangerous ``forward'' and ``asymmetric'' regions of phase space in the
lab frame.

As an important reference, we will start by recalling the kinematic properties of $A'$ production
and decay using the results of Appendix \ref{app:Details}.  We again
consider a monochromatic incident electron beam of energy $E_0$.  
Let $\theta_{cm}$ be the emission angle of the forward decay product relative to the $A'$ 
direction in the $A'$ rest frame. Let $\theta_{A'}$ be the emission angle of the $A'$ relative to the 
beam direction in the lab frame. 
As we have shown, the characteristic $A'$ emission angle is small and
is set by 
$\theta_{A'} \sim \f{m_{A'}\sqrt{\overline{1-x}}}{E_0}$, where $x\equiv E_{A'}/E_0$. 
In the limit of small $\f{m_{A'}}{xE_0}$,
the lab frame opening angles $\theta_{\pm}$ and energies $E_{\pm}$ of the $A'$ 
decay products are,
\bea
E_{\pm} &=& \f{xE_0}{2}(1\pm \cos(\theta_{cm})) \label{eq:DecayEnergy},\\
\tan(\theta_{\pm}) &=& \pm \f{1}{\gamma} \sqrt{\f{1\mp \cos(\theta_{cm})}{1\pm \cos(\theta_{cm})}} + \tan(\theta_{A'})
\label{eq:DecayAngle},
\eea
where $\gamma=\f{xE_0}{m_{A'}}$. 

The characteristic transverse momentum of the $A'$ is
 $p_{A',\perp}\approx E_{A'}\theta_{A'}\sim \sqrt{1-x}\,m_{A'}$,
while the typical recoil of the target is $|q_{min}|\approx
\left(\f{m_{A'}}{2 xE_0}\right)m_{A'}$.  
The median value of $1-x$ is $\overline{1-x}\approx \f{m_{A'}}{E_0}$ for $m_{A'}\gsim 50 \MeV$,
implying that $|q_{min}|$ is parametrically smaller than $p_{A',\perp}$ by $\sim \sqrt{\f{m_{A'}}{E_0}}$. 
Evidently, the recoiling electron largely balances the recoil of the $A'$.  
 The energy $E_{R}$ and angle $\theta_{R}$ of the recoiling final state beam electron in the 
 laboratory frame is,
 \bea
 E_{R} &=& (1-x)E_0 \approx m_{A'} \label{eq:RecoilEnergy}, \\
 \tan(\theta_{R}) &\approx& \sqrt{\f{m_{A'}}{E_0}}(1+\f{m_{A'}}{2E_0}+...) \label{eq:RecoilAngle}. 
 \eea
Note the relatively wide angle of the recoiling electron relative to the $A'$ decay products. 
Equations (\ref{eq:DecayEnergy})--(\ref{eq:RecoilAngle}) summarize 
the important kinematic characteristics of $A'$ production. 

\begin{figure}
\halfPage{\includegraphics[width=0.9\textwidth]{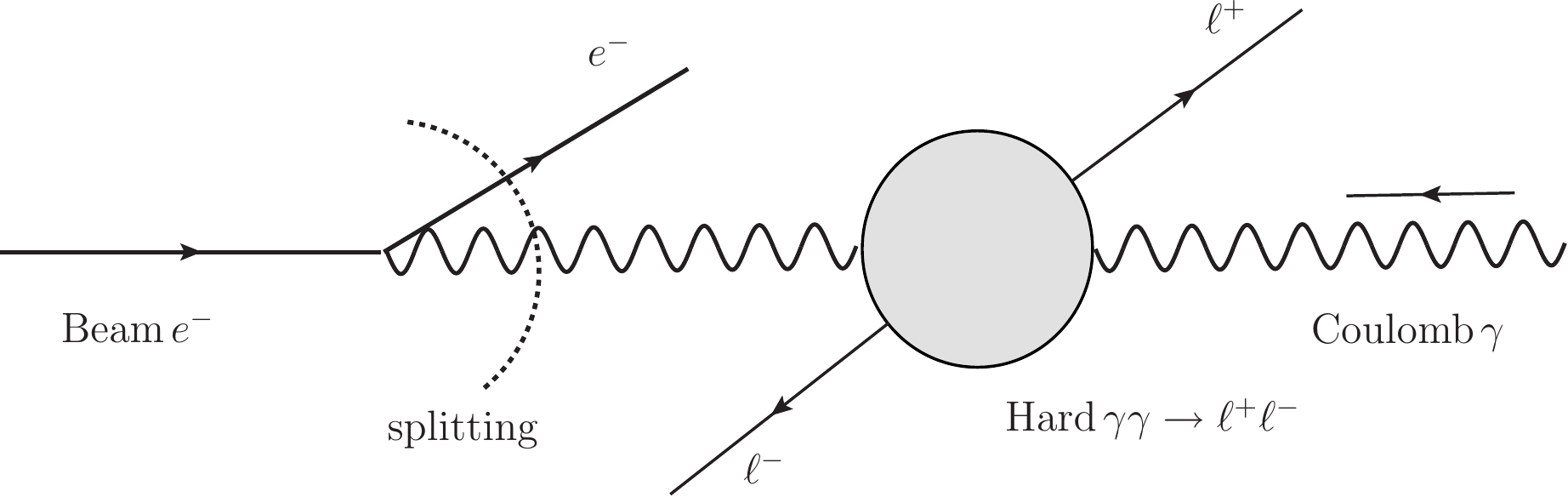}}
\caption{Bethe-Heitler reactions viewed as hard $\gamma\gamma\rightarrow l^+l^-$
processes.\label{fig:BH1}}
\end{figure}

As with $A'$ production, trident reactions can also be analyzed using
the Weizs\"acker-Williams approximation, where we group an outgoing
$\ell^+\ell^-$ pair with fixed invariant mass $m^2$ to act as the $A'$
candidate.  In the case of Bethe-Heitler production, we can further
approximate the beam electron as splitting at small angle into the
recoil electron and a nearly on-shell photon, which scatters with the
Coulomb photon into the $\ell^+\ell^-$ pair (see Figure \ref{fig:BH1}).
There are two sources of large logs in the Bethe-Heitler
cross-section: the soft and collinear logs in the photon radiation by
the electron and a forward-scattering log in the
$\gamma\gamma\rightarrow \ell^+\ell^-$ subprocess.  Both are regulated
by $m_e$, and can be further suppressed by kinematic requirements.  We
postpone the derivation --- a matter of standard results and kinematic
bookkeeping --- to the following section, and focus here on its
implications.

In this approximation, upon integrating over an invariant mass window
of size $\delta m$ about $m_{A'}$, we find 
\bea 
\f{d\sigma}{d x
  d\cos\theta_{A'} d\hat c}= &&
\frac{2 \alpha^4\chi}{\pi} 
\f{\delta m}{m_{A'}}\f{1}{m_{A'}^2}
 \frac{1+(1-x)^2}{\theta_{A'}^2 x} \nonumber \\
&&\qquad \times \left(\f{1+\hat c}{1-\hat c}+ \f{1-\hat c}{1+\hat c}\right), 
\label{eq:bhXSEC}
\eea 
where
$x$ and $\theta_{A'}$ are defined as before, and $\hat c =
\cos\theta_{cm}$, where $\theta_{cm}$ is measured relative to the axis
of the incoming photons in the $\gamma\gamma \rightarrow \ell^+\ell^-$
process, which is near enough to the beam axis for our purposes.  
This displays the expected singularities at small $x$ (soft), $|\hat c|
\rightarrow 1$ (forward scattering), and small $\theta_{A'}$
(collinear).  This is quite different behavior from the signal and
radiative backgrounds, which are peaked at \emph{large} values of $x$
and slowly varying in $\hat c$.

Requiring $x$ above $1-\delta$, with $\delta$ near or below its median
value $\bar \delta = \max(m_{A'}/E_0, m_e/m_{A'})$ keeps a large
fraction of the signal and suppresses the Bethe-Heitler background by
a factor of $\delta$.  Likewise, the signal is relatively flat in
$\hat{c}=\cos\theta_{cm}$.  According to \eqref{eq:DecayEnergy} and
\eqref{eq:DecayAngle}, we can ensure modest $\hat c$ by constraining
the ratio of the lab-frame energies or of the opening angles of the
two decay products to be near unity.  

There remains the collinear singularity at small $\theta_{A'}$.  The
signal is also peaked forward, but with a singularity regulated by
$m_{A'}$ rather than $m_e$, so that it is produced at much wider
characteristic $A'$ angles than Bethe-Heitler processes.  Since the
absolute angles $\sim (m_{A'}/E_0)^{3/2}$ are still small and must be
obtained by summing two momenta, it is impractical to place a lower
bound on the $A'$ emission angle. It is probably much easier to cut on
the angle $\theta_R$ of the recoiling electron.  Since the recoiling electron has
much lower energy at large $x$ and approximately balances the
transverse momentum of the $A'$, it is emitted at a much larger angle.
For the median $x$, the electron energy $\sim m_{A'}$ and its angle
$\sim (m_{A'}/E_0)^{1/2}$.  Requiring the recoiling electron momentum 
and angle near these values significantly reduces the Bethe-Heitler rate.  
In a similar spirit, it
may or may not also be easier to implement a tight $x$ cut using the
energy of the recoil electron rather than the total energy of the
decay products making up the $A'$ candidate.  After these cuts, we
find
\bea
&&\sigma_{cut}(x_{min},\theta_{R,min},|\hat c|_{max})
\approx \frac{16\alpha^4\chi}{\pi}\f{\delta m}{m_{A'}}\f{1}{m_{A'}^2}
\nonumber \\
&&\qquad \times \log\left((1-x_{min})^{-1} \theta_{R,min}^{-1}\right)
(1-x_{min})\nonumber \\ %}\log(x_{min}^{-1})\nonumber \\
&&\qquad \times \left(\tanh^{-1} |\hat c|_{max}  -|\hat c|_{max}/2\right).\label{bhref}
\eea
When $\theta_{cut}$ is small relative to the typical angular spread
$\f{m_{A'}}{E_0} (1-x)$ of the signal and radiative backgrounds, the
factor on the first line is related to the lepton-pair cross-section
$\sigma_{rad}$ with the same cuts from the radiative diagrams alone by
\be
9 \sigma_{rad} \left( \log \f{1-x_{min}}{1-x_{max}}\right)^{-1}
\epsilon_{rad}^{-1}(\theta_{R,min}, |\hat c|_{max}),
\ee
where $1-x_{max} = \max(m_A^2/E_0^2, m_e^2/m_A^2)$ is the value of $x$
where the log divergence in $\sigma_{rad}$ is regulated, and
$\epsilon_{rad}$ the efficiency for the radiative process (or signal)
to pass $\theta_{R}$ and $|\hat c|$ selections.

Let us consider a representative case, $E_0=5$ GeV, $m_{A'} = 0.5$
GeV. Requiring $x>0.9$, $\theta_R > 1/10$, and $E_+/E_- <3$
(i.e. $\hat c < 0.5$), we retain roughly 20\% of the signal and reduce
the contribution of Bethe-Heitler to the signal region to roughly the
same size as the radiative contribution.  A proper optimization of
these cuts is best done with full monte carlo for the background, and
of course depends on the characteristics of an individual experiment,
but we have confirmed numerically that the kinematic differences
between Bethe-Heitler and radiative production are sufficient that it
can be made sub-dominant while maintaining high efficiency for the
$A'$ signal.

In the case of $A'\rightarrow e^+e^-$, the Bethe-Heitler process
can also contribute with the electron labeled $\ell^-$ in Figure
\ref{fig:SigAndBkg} identified as the recoiler and $\ell^+$ and $e^-$ forming
the $A'$ candidate.  Here the recoil electron kinematics is as in the signal
process but there is a forward-scattering singularity when most of the
$A'$-candidate energy is carried by the $e^-$ with a softer $e^+$.
The cuts above remove this singularity as well, and we will not discuss
it further.

\subsection*{Bethe-Heitler Pair Production in the Collinear Splitting
  Approximation }
We now extend the earlier Weizsacker-Williams treatment to compute the Bethe-Heitler pair production cross-section.
Using the notation defined in Appendix \ref{app:Details}, but now with an outgoing lepton/anti-lepton
pair with momenta $l^-$/$l^+$, the fully differential cross section is,
\bea\label{eq:BkgCrossSection}
&&\f{d\sigma(p_1 + P_i \rightarrow p_2 + l^++l^-+P_f}{d E_{A'}
  d\cos\theta_{A'}dm^2 d\hat{t} d\phi_d}=
\left(\f{\alpha \chi}{\pi}\right) \left(\f{E_0 x \beta_{A'}}{(1-x)}\right) \qquad \nonumber \\
  && \qquad\qquad \ \ \times \f{d\sigma(p_1 + q \rightarrow p_2 + l^++l^-)}{d(p_1\cdot
  a)  dm^2 d\hat{t} d \phi_d}\bigg|_{t=t_{min}} \label{eq:WWbkg}
\eea
where $a=l^++l^-$ is the total four-momentum of the $A'$ candidate,
$m^2=a^2$ their invariant mass, and
$\hat{t}=(l^+-q)^2=(l^+-P_f+P_i)^2$.  $\phi_d$ the angle between $p_1$
and $l^+$ in the rest frame of $a$. As before, the hard sub-processes
($2\rightarrow 3$ in this case) are separated from the soft Coulomb
photon exchange.

Starting from equation (\ref{eq:BkgCrossSection}), we will analyze the behavior of
Bethe-Heitler reactions (see Figure \ref{fig:SigAndBkg} (b)) relative
to $A'$ production in a leading logarithm approximation.   This
suffices to identify the singularities.

% %
%% \begin{figure}
%% \halfPage{\includegraphics[width=0.9\textwidth]{figures/BH2.pdf}}
%% \caption{\textbf{[N wants to replace]}$A'$ production and $\gamma^*$ backgrounds for $l^+l^-$ viewed as
%% hard $e^+e^-\rightarrow l^+l^-$ processes.\label{fig:BH2}}
%% \end{figure} 
%% %

To write the right-hand side of \eqref{eq:BkgCrossSection} in the approximation
of near-collinear splitting, it is useful to introduce the more
familiar $z$ and $p_\perp$ variables and relate them to the kinematic variables in the 
center-of-mass frame of the beam electron and Coulomb photon 
and to the lab-frame.  We recall that in the lab frame,
the Coulomb photon is purely spacelike, with $|\vec q| =\sqrt{t_{min}}
= \frac{U(x,\theta_{A'})}{2 E_0 (1-x)} \approx m^2/(2E_0x)$ for small $\theta_{A'}$
(we have replaced $m_{A'}^2$ with the invariant mass $m^2$ for the
off-shell processes, but the kinematic conditions are unchanged).  
As $|\vec q|$ at $t_{min}$ kinematics depends on $x$ and $\theta_{A'}$, the
CM frame does too --- it is obtained from the  CM frame by boosting
with $\beta \approx 1-q/E_0$ (we drop much smaller $\theta_{A'}$-dependent
corrections) , leading to CM-frame momenta for the incoming electron
and Coulomb photon, 
\bea 
p_1^{cm} & =& (p,0,0,p)\\
q^{cm}   & =& (\beta p,0,0,-p)
\eea
where $p = \sqrt{\frac{qE}{2(1-q/E)}}$.  The recoil electron has
momentum 
\bea
p_2^{lab}& =& \bigg((1-x) E_0, E_0 \theta_{A'} x, 0, \nonumber \\
& & \qquad \qquad (1-x) E_0 - 
  \frac{E_0 \theta_{A'}^2 x^2}{2 (1-x)}\bigg),\\
p_2^{cm} &= &(z p, p_\perp, 0, \sqrt{z^2  p^2 - p_\perp^2}),
\eea
where 
\be
z \equiv 1-x, \quad p_\perp \equiv E_0 \theta_{A'} x,
\ee
up to corrections of higher order in  $\theta_{A'}$ and $m_e$.  For
small $p_\perp$, we can now approximate 
\be
\f{d\sigma}{dz dp_\perp^2 d\hat t  d\phi_d} = 
\left ( \f{\alpha}{2\pi} \right ) \f{[ 1 + z^2 ]}{(1-z)p_{\perp}^2}
\f{d \sigma(\gamma\gamma\rightarrow l^+l^-)}{d\hat{t} d\phi_d},
\ee
where 
\bea
\f{d \sigma(\gamma\gamma\rightarrow l^+l^-)}{d\hat{t} d\phi_d} &&=
\f{\alpha^2}{m^4} \left( \frac{\hat t}{\hat u} + \frac{\hat
  u}{\hat t}\right).
\eea
We emphasize that the recoil electron is always right-moving, so
the splitting approximation is valid where the Bethe-Heitler
cross-section is largest, but is \emph{never} a good approximation for
the radiative (or signal) processes, where the $\gamma^*\rightarrow
e^- e^{+*}$ ``splitting'' always produces an electron going backwards
relative to the $\gamma^*$.  

The kinematic variables of \eqref{eq:BkgCrossSection} are related to
$z, p_\perp, \phi_d$ by 
\bea
m^2 &= &a^2 = (1+\beta - 2 z)(1+\beta) p^2 \\
p_1.a &=& (1+\beta-z)p^2 +z p^2 \sqrt{1-\frac{p_\perp^2}{z^2 p^2}}\\
\phi_d &=& \phi_d, \label{changeVars3Body}
\eea
leading to a Jacobian factor 
\be
\frac{d^2}{d(p_1.a) dm^2} \approx \frac{z}{p^2 (1+\beta)}
\frac{d^2}{dp_\perp^2  dz} \approx  \frac{2 x (1-x)}{m^2}
\frac{d^2}{dp_\perp^2  dz}.
\ee
Taking $1+\beta \rightarrow 2$, $p^2 \rightarrow \frac{|\vec q| E_0}{2} \approx
\frac{m^2}{4 x}$, we find
\begin{widetext}
\bea
\f{d\sigma(p_1 + q \rightarrow p_2 + l^++l^-)}{d(p_1\cdot
  a)  dm^2 d\hat{t} d \phi_d}\bigg|_{t=t_{min}(x,\theta_{A'})}
 & =  & 
\frac{\alpha}{2\pi} \frac{1+(1-x)^2}{E_0^2 \theta_{A'}^2 x^3} % prefactor
\frac{2 x(1-x)}{m^2}                                      % jacobian
\f{\alpha^2}{m^4} \left(\f{\hat t}{\hat u}+ \f{\hat u}{\hat
  t}\right), %2->2 process
\eea
and hence by \eqref{eq:WWbkg},
\be
\f{d\sigma}{d x
  d\cos\theta_{A'}dm^2 d\hat{t} d\phi_d}=
\frac{\alpha^4\chi}{\pi^2} 
\frac{1+(1-x)^2}{\theta_{A'}^2 x} % prefactor
\f{1}{m^6} \left(\f{\hat t}{\hat u}+ \f{\hat u}{\hat
  t}\right).
\ee
\end{widetext}
Changing variables from $\hat t$ to $\hat c = \cos\theta_{cm}$ of the
$2\rightarrow 2$ process, $\hat t = \f{m^2}{2}(1-\hat c)$, and
integrating over $\phi_d$ and over $m^2$ from $m_{A'}^2-m_{A'}\delta m$ to $m_{A'}^2 +
m_{A'}\delta m$, we obtain the result of Equation \eqref{eq:bhXSEC}.

%%%%%%%%%%%%%%%%%%%%%%%%%%%%%%%%%%%
%%%%%%%%%%%%%%%%%%%%%%%%%%%%%
\bibliographystyle{apsrev}
\bibliography{NewFixedTargetExp}
\end{document}